\documentclass[a4paper,11pt]{article}

\usepackage{float}
\usepackage{jheppub}

\usepackage{empheq,mathtools}
\usepackage{amsmath,amsfonts,amssymb}
\usepackage{mathrsfs}
\usepackage{cancel,slashed}
\usepackage{tensor}
\usepackage{xcolor}
\usepackage[normalem]{ulem}
\usepackage{graphicx}
\usepackage{subfigure}

\makeatletter
\def\@fpheader{\relax}
\makeatother

\newcommand{\be}{\begin{equation}}
\newcommand{\ee}{\end{equation}}

\usepackage[czech,english]{babel}
\usepackage{yfonts}

\usepackage{url}
\usepackage{booktabs,tabularx,placeins}
\usepackage{mathabx}

\preprint{{USTC-ICTS/PCFT-26-90}}

\usepackage{tikz}
\usetikzlibrary{positioning, arrows.meta}

\title{\boldmath Eight-Charge Rotating AdS Black Holes in Gauged STU Supergravity}

\affiliation[a]{Interdisciplinary Center for Theoretical Study, University of Science and Technology of China, Hefei 230026, China}
\affiliation[b]{Peng Huanwu Center for Fundamental Theory, Hefei 230026, China}
\affiliation[c]{School of Physics and Shing-Tung Yau Center, Southeast University, Nanjing, 210096, China}

\author[a,b]{Yuan-Tai Wang}
\author[c]{Wenni Zheng}

\emailAdd{wangyuantai@ustc.edu.cn}
\emailAdd{wennizheng@seu.edu.cn}

\abstract{
We construct a family of exact non-extremal rotating AdS black hole solutions carrying four electric and four magnetic charges in four-dimensional gauged STU supergravity.
We establish the local independence of all eight electromagnetic charges, including on a non-extremal branch at fixed spindle orders, with boundary electromagnetic sources allowed to vary.
This branch has a regular exterior away from the prescribed orbifold loci.
Within this family, smoothness of both axes imposes a nontrivial quartic charge relation, leaving at most seven independent electromagnetic charges.
Generic eight-charge solutions carry an axial conical defect and have non-conformally-flat boundary geometries.
We determine the horizon quantities and extremal near-horizon geometries and matter fields, give sufficient conditions excluding closed timelike curves in the exterior away from the defect core, and exhibit an extremal spindle branch.
}

\begin{document} 
\maketitle
\flushbottom

\section{Introduction}\label{sec:intro}

Four-dimensional gauged STU supergravity, with four Abelian gauge
fields and three complex scalars, admits an embedding in
eleven-dimensional supergravity and provides a tractable setting for
charged AdS black holes and their holographic
interpretation~\cite{Maldacena:1997re,Cvetic:1999xp,Azizi:2016embedding}.
In the \(\mathrm{AdS}_4\times S^7\) embedding, the gauge fields correspond
to the \(U(1)^4\) Cartan subgroup of the \(SO(8)\) isometry of \(S^7\).
Their electric charges are internal angular momenta and are identified
with the Cartan charges of the R-symmetry in the boundary
theory~\cite{Cvetic:1999xp}.
The global interpretation of exact solutions requires an analysis
of their horizons and causal structure~\cite{Stephani2003}.
The problem is to realize independently variable electric and magnetic
charges in a non-extremal rotating family, and to determine how this
freedom is restricted by axial regularity and AdS boundary data.

In the dual ABJM theory~\cite{Aharony:2008ug}, the microscopic
entropy of supersymmetric rotating electrically charged AdS\(_4\)
black holes has been reproduced in the large-\(N\) Cardy
limit~\cite{Nian:2019pxj,Choi:2019zpz}. For non-extremal rotating
electrically charged black holes, boundary calculations in a
matrix-model approximation find qualitative agreement with supergravity
in the high-temperature regime~\cite{Nian:2025iei}.

General non-extremal rotating families provide higher-dimensional
benchmarks, with three electric charges and two independent rotations
in five-dimensional gauged STU supergravity~\cite{Wu:2012general5d}
and two electric charges and three independent rotations in
seven-dimensional \(U(1)^2\) gauged supergravity~\cite{Bobev:2023ads7}.
In five dimensions, supersymmetric
solutions~\cite{Gutowski:2004ez,Gutowski:2004yv,Kunduri:2006ek}
have motivated studies of BPS states, superconformal indices and
microscopic entropy in the dual \(\mathcal N=4\) gauge
theory~\cite{Grant:2008sk,Chang:2013fba,Cabo-Bizet:2018ehj,Choi:2018hmj,Benini:2018mlo,Cabo-Bizet:2019eaf,Nian:2020qsk,Murthy:2020scj,Agarwal:2020zwm,Chang:2024zqi}.
In seven dimensions, entropy extremization and microscopic
calculations use the six-dimensional \(\mathcal N=(2,0)\) index and
five-dimensional maximally supersymmetric Yang--Mills
theory~\cite{Hosseini:2018dob,Choi:2018hmj,Kantor:2019lfo,Nahmgoong:2019hko,Bobev:2025ads7holography}.
Representative solutions are summarized in
Appendix~\ref{app:higher-dimensional-comparison}.
These higher-dimensional families provide a comparison for the
four-dimensional problem, where a Maxwell field strength and its
electromagnetic dual are both two-forms and the asymptotic two-sphere
supports magnetic monopole flux. Each of the four STU vectors can
therefore carry both electric and magnetic charges, and the theory
admits the electromagnetic duality rotations used below.

The static sector already accommodates a broad range of
electromagnetic charges and horizon topologies. Non-extremal
AdS black holes with four electric charges were constructed in the
\(U(1)^4\) truncation of \(SO(8)\) gauged
supergravity~\cite{Duff:1999electric}, together with their
higher-dimensional interpretation~\cite{Cvetic:1999xp}. Early electric
BPS configurations~\cite{Sabra:1999bps} and magnetic and dyonic
solutions~\cite{Caldarelli:1999bps,Chamseddine:2000dyonic} established
the importance of supersymmetry and horizon topology. Regular static
supersymmetric black holes with nonconstant scalars were subsequently
obtained~\cite{Cacciatori:2010bps}, followed by more general dyonic
BPS families~\cite{Katmadas:2014static,Halmagyi:2015dyonic}.
Microscopic entropy counting for static magnetic and dyonic black holes
was subsequently achieved in the dual field
theory~\cite{Benini:2015eyy,Benini:2016rke}.\footnote{Ref.~\cite{Benini:2015eyy} counts the magnetic STU black
holes constructed in Ref.~\cite{Cacciatori:2010bps} and further studied
in Ref.~\cite{Hristov:2011static}. Ref.~\cite{Benini:2016rke} extends
the counting to dyonic black holes in the static BPS families of
Refs.~\cite{Katmadas:2014static,Halmagyi:2015dyonic}.}

Away from supersymmetry, magnetic and dyonic solutions were
constructed in STU
truncations~\cite{Toldo:2012nonextremal,Gnecchi:2013flow}, and the
real formulation of special geometry yielded further families,
including black holes with four magnetic
charges~\cite{Klemm:2013realI,Klemm:2013realII}. The thermodynamics of
static dyonic solutions also revealed the role of scalar hair and
boundary conditions~\cite{Lu:2013dyonic}. Analytic neutral hyperbolic
black holes with scalar hair can arise as limits of charged solutions
in theories of gravity coupled to an Abelian field and a dilaton,
including special truncations of gauged STU
supergravity~\cite{Ren:2019lgw}. The general non-extremal static STU
family carries four independent electric and four independent
magnetic charges and allows spherical, planar and hyperbolic
horizons~\cite{Chow:2014dyonic}.

Rotation introduces angular dependence in the axions and dilatons,
and the available non-extremal AdS solutions initially covered
restricted charge sectors. These included pairwise equal charges
and their NUT extensions~\cite{Chong:2004na}, with thermodynamics,
supersymmetric limits and global regularity studied
in~\cite{Cvetic:2005zi}, as well as single-charge and two-charge
solutions with NUT
generalisations~\cite{Chow:2011single,Chow:2011two}. Rotating dyonic
families with nonconstant scalars were also constructed in the
single-vector-multiplet
truncation~\cite{Chow:2014dyonic,Gnecchi:2014rotating}. In particular,
the complete solution with two electric charges, two magnetic charges
and an independent NUT parameter identifies the four STU gauge
fields in pairs~\cite{Chow:2014dyonic}. It therefore has fewer
independent charges than the general static family.

Acceleration and more general horizon geometries provide further
extensions. In gravity coupled to a single Abelian field with a
cosmological constant, rotation, NUT charge, acceleration and dyonic
charges can be combined in a single
family~\cite{Plebanski:1976gy,Griffiths:2006pd}. This family embeds in
the minimal STU truncation with identified gauge fields and constant
scalars. Beyond that truncation, static accelerating solutions with
four electric charge parameters and nontrivial dilatons were
constructed~\cite{Lu:2014sza}. Supersymmetric rotating and
accelerating solutions in the minimal theory admit spindle horizons
and smooth eleven-dimensional uplifts~\cite{Ferrero:2020twa}.
Multicharge extensions include complete nonrotating accelerating
black holes and rotating supersymmetric spindle near-horizon
geometries~\cite{Ferrero:2021ovq}. The latter do not by themselves
specify complete rotating black-hole solutions.

Supersymmetry supplies a separate route to rotating solutions.
Rotating BPS black holes with a nonconstant scalar were constructed
in a single-vector-multiplet model~\cite{Klemm:2011rotating}.
More general rotating attractors and black holes with a magnetic
topological twist~\cite{Hristov:2019attractors} were followed by
matter-coupled rotating supersymmetric AdS black holes, including
explicit dyonic STU solutions~\cite{Hristov:2019dyonic}. These
zero-temperature families obey supersymmetry and regularity
constraints and do not determine the general non-extremal solution.
The scalar potential introduced by gauging generically breaks the
hidden symmetries that underlie solution-generating methods in ungauged
supergravity, leaving a restricted set of
transformations~\cite{Cacciatori:2016duality}. This obstructs a direct
extension of the ungauged construction to generic rotating dyonic AdS
black holes. Specific compact electromagnetic duality transformations
still preserve the bosonic equations, whereas preserving supersymmetry
also requires compatibility with the fixed gauging
vector~\cite{Halmagyi:2014duality,Cacciatori:2016duality}. The
corresponding restrictions must therefore be distinguished when
counting non-extremal and BPS charge parameters.

In this paper, we construct a non-extremal rotating family with eight
locally independent classical electromagnetic charges. We give the
complete metric, six real scalars, four gauge potentials and their
four local Maxwell duals as explicit functions of the coordinates and
parameters.
We determine the horizon quantities of the eight-charge family,
including its entropy, temperature, angular velocity and electromagnetic
potentials. We also derive the extremal near-horizon metric and matter
fields and obtain sufficient conditions excluding closed timelike
curves in the exterior.
The generic eight-charge branch carries boundary electromagnetic
sources and an axial conical defect. Its complete thermodynamic
description remains open. The \(\gamma=0\) limit yields a complete
rotating seven-charge black-hole solution with smooth axes.
Setting its three compact duality angles to zero further recovers the
full rotating AdS black-hole solution with four independent \(U(1)\)
charge parameters recently obtained in
Refs.~\cite{Ma:2026fourcharge,Wu:2026fourcharge}.
For the seven-charge solution, we give seed-normalized thermodynamic
quantities and verify a first-law identity on the solution parameter
space. We also determine the BPS limit and analyze the absence of
closed timelike curves. 
We show that the supersymmetric
entropy function reproduces the Bekenstein--Hawking entropy on the
established BPS branch. Table~\ref{tab:ads4-solutions} summarizes the
four-dimensional solutions reviewed above and the family constructed
here.

\begin{table}[!htbp]
\centering
\small
\setlength{\tabcolsep}{7pt}
\setlength{\arrayrulewidth}{0.4pt}
\renewcommand{\arraystretch}{1.45}
\caption{Exact AdS\(_4\) black holes in gauged STU supergravity and its
truncations.
}
\label{tab:ads4-solutions}
\begin{tabularx}{\textwidth}{|>{\raggedright\arraybackslash}X|>{\centering\arraybackslash}p{0.23\textwidth}|}
\hline
Black hole parameters & References \\
\hline
Static electric: \(Q_I\)
& \cite{Duff:1999electric} \\
\hline
Static dyonic: \(Q_I,P^I\)
& \cite{Chow:2014dyonic} \\
\hline
Static BPS: \(Q_I,P^I\)
& \cite{Cacciatori:2010bps,Katmadas:2014static,Halmagyi:2015dyonic} \\
\hline
Pairwise electric: \(Q_1=Q_3,\ Q_2=Q_4;\ a,n\)
& \cite{Chong:2004na} \\
\hline
Single electric charge: \(Q_1;\ a,n\)
& \cite{Chow:2011single} \\
\hline
Two electric charges: \(Q_1,Q_2;\ a,n\)
& \cite{Chow:2011two} \\
\hline
Dyonic \(U(1)^2\) truncation: \(Q_I,P^I;\ a,n\)
& \cite{Chow:2014dyonic} \\
\hline
Rotating BPS: \(Q_I,P^I;\ a\)
& \cite{Hristov:2019attractors,Hristov:2019dyonic} \\
\hline
Electric C-metric: \(Q_I;\ A_{\mathrm{acc}}\)
& \cite{Lu:2014sza} \\
\hline
Four-charge family: \(Q_I;\ a\)
& \cite{Ma:2026fourcharge,Wu:2026fourcharge} \\
\hline
Seven-charge limit: \(Q_I,P^I;\ a\)
& This work \\
\hline
Eight-charge branch: \(Q_I,P^I;\ a\)
& This work \\
\hline
\end{tabularx}
\par\vspace{4pt}
\begin{minipage}{\textwidth}
\footnotesize
\textbf{Notation}: \(Q_I\) and \(P^I\) denote electric and magnetic
charges, and \(a\) is the rotation parameter, as in
section~\ref{sec:seven-charge}. Here \(I=1,\ldots,4\), except
\(I=1,2\) in the \(U(1)^2\) row, where the gauge fields are
obtained by pairwise identification. The \(n\)-entries refer to the
earlier NUT extensions, and \(A_{\mathrm{acc}}\) to acceleration.
Charge labels allow relabelling and refer to the indicated duality frame. A mass parameter is implicit outside the BPS rows.
The static dyonic family admits spherical, planar and hyperbolic
horizons. The recent four-charge rotating family is expressed in the
electric frame used here. Its original mixed frame has two electric
and two magnetic charges.
BPS data obey the applicable twist, supersymmetry and regularity
constraints. For the last two rows, the charge constraints and
boundary data are specified in sections~\ref{sec:eight-global} and
\ref{sec:seven-subfamily}. The eight-charge count allows boundary
sources and axial defect data to vary.
\end{minipage}
\end{table}

The paper is organized as follows. 
Section~\ref{sec:seven-charge} presents the eight-charge solution,
including its charges, boundary data and seven-charge limit.
Section~\ref{sec:eight-properties} discusses the
general horizon quantities, extremal near-horizon fields and sufficient
causality conditions, together with the restrictions needed for a
supersymmetric interpretation. Section~\ref{sec:discussion} discusses
the results and open problems. Appendix~\ref{app:stu-eom} collects the
Lagrangian, field equations, duality conventions and auxiliary
functions. Appendix~\ref{app:seven-subfamily} gives detailed
thermodynamic and BPS results for the seven-charge subfamily, including
its supersymmetric entropy function.
Appendix~\ref{app:eight-spindle-example} gives extremal and
non-extremal spindle examples within the eight-charge family, and
exhibits a regular non-extremal branch with rank-eight physical fluxes
at fixed spindle orders.
Appendix~\ref{app:Phase space formalism} presents the horizon symmetry
and surface-charge analysis, together with the assumptions underlying
the Cardy entropy matching. Appendix~\ref{app:higher-dimensional-comparison}
collects the higher-dimensional comparison tables.

\FloatBarrier

\section{Rotating AdS black holes with eight electromagnetic charges}
\label{sec:seven-charge}

Four-dimensional gauged STU supergravity is an \(\mathcal N=2\)
theory coupled to three vector multiplets. It arises as the
\(\mathrm U(1)^4\) Cartan truncation of \(\mathcal N=8\),
\(\mathrm{SO}(8)\) gauged supergravity and has scalar manifold
\([\mathrm{SL}(2,\mathbb R)/\mathrm{SO}(2)]^3\)
\cite{Cvetic:1999xp,Azizi:2016embedding}. Further consistent
truncations follow by identifying gauge fields and restricting the
scalars. In an electric frame, the \(2+2\) and \(3+1\)
identifications give the \(X^0X^1\) and \(T^3\) models,
respectively, each with two gauge fields and one complex scalar
\cite{Azizi:2016embedding,Hristov:2019dyonic}. Identifying all four
gauge fields and fixing the scalars at the AdS vacuum yields minimal
\(\mathcal N=2\) gauged supergravity, whose bosonic sector is
Einstein--Maxwell theory with a negative cosmological
constant~\cite{Chow:2014dyonic}.

The equal-coupling bosonic equations admit a compact
\(\mathrm{SO}(2)^3\) electromagnetic duality symmetry. The connected
stabilizer of the fixed Fayet--Iliopoulos (FI) vector is the
\(\mathrm{SO}(2)^2\) subgroup relevant to the BPS sector
\cite{Cacciatori:2016duality}.

In this section, we present the rotating solution with eight independent
electromagnetic charges. We first specify the supergravity conventions
and then give the complete metric and matter fields, followed by their
electromagnetic charges, admissible black-hole branches and
boundary conditions, and the seven-charge subfamily.

\subsection{The theory}
\label{sec:seven-theory}

The bosonic sector of gauged STU supergravity consists of the metric,
four Abelian vectors and three dilaton--axion pairs
\((\varphi_i,\chi_i)\), with \(i=1,2,3\). Gauging introduces a scalar
potential that admits an AdS vacuum. To specify the Lagrangian and
its duality frame, we follow the conventions of
Ref.~\cite{Chong:2004na}.\footnote{We use
signature \((-+++)\) and orientation
\(\epsilon_{tr\theta\phi}=\sqrt{-\det(g_{\mu\nu})}\), equivalently
\(\epsilon_{t_0r\sigma u}=\sqrt{-\det(g_{\mu\nu})}\) in Carter coordinates.
For a \(p\)-form
\(\mathcal X_{(p)}=\frac1{p!}\mathcal X_{\nu_1\cdots\nu_p}
\mathrm dx^{\nu_1}\wedge\cdots\wedge\mathrm dx^{\nu_p}\),
with \(0\leq p\leq4\), the Hodge dual is
\[
 (\star\mathcal X_{(p)})_{\mu_1\cdots\mu_{4-p}}
 =\frac1{p!}\epsilon_{\mu_1\cdots\mu_{4-p}\nu_1\cdots\nu_p}
                      \mathcal X^{\nu_1\cdots\nu_p}\,.
\]
The alternating symbol is normalized by
\(\widetilde\epsilon^{t_0r\sigma u}=+1\).}
The scalar potential and complex coordinates are
\begin{equation}
 V=-g^2\sum_{i=1}^3
       \left(2\cosh\varphi_i+\chi_i^2e^{\varphi_i}\right)\,,
 \qquad z_i=\chi_i+\mathrm{i}e^{-\varphi_i}\,.
\label{eq:fourcharge-potential}
\end{equation}
The coupling \(g>0\) fixes the AdS vacuum, with
\(\Lambda=-3g^2\). We use the mixed
electric--magnetic ordering \(A^1=A_{(1)1}\), \(A^2=A_{(1)2}\),
\(A^3=\mathcal A_{(1)}^1\) and \(A^4=\mathcal A_{(1)}^2\), with
\(f^I=\mathrm dA^I\), \(I,J=1,\ldots,4\).
The bosonic Lagrangian is given by
\begin{equation}
\begin{aligned}
 \mathcal L_4={}&R\star 1
 -\frac12\sum_{i=1}^3\star \mathrm{d}\varphi_i\wedge\mathrm{d}\varphi_i
 -\frac12\sum_{i=1}^3e^{2\varphi_i}
                         \star \mathrm{d}\chi_i\wedge\mathrm{d}\chi_i\\
 &-\frac12e^{-\varphi_1}\Big(
 e^{\varphi_2-\varphi_3}\star F_{(2)1}\wedge F_{(2)1}
 +e^{\varphi_2+\varphi_3}\star F_{(2)2}\wedge F_{(2)2}\\
 &\hspace{34mm}
 +e^{-\varphi_2+\varphi_3}\star \mathcal F_{(2)}^1\wedge\mathcal F_{(2)}^1
 +e^{-\varphi_2-\varphi_3}\star \mathcal F_{(2)}^2\wedge\mathcal F_{(2)}^2\Big)\\
 &-\chi_1\left(F_{(2)1}\wedge\mathcal F_{(2)}^1
                   +F_{(2)2}\wedge\mathcal F_{(2)}^2\right)-V\star 1\,,
\end{aligned}
\label{eq:fourcharge-lagrangian}
\end{equation}
where the axion-dependent field-strength combinations are 
\begin{equation}
\begin{aligned}
 F_{(2)1}&=f^1-\chi_2f^4\,,\\
 \mathcal F_{(2)}^1&=f^3+\chi_3f^4\,,\\
 F_{(2)2}&=f^2+\chi_2f^3-\chi_3f^1+\chi_2\chi_3f^4\,,\\
 \mathcal F_{(2)}^2&=f^4\,.
\end{aligned}
\label{eq:fourcharge-dressed-strengths}
\end{equation}
Dualizing vectors 1 and 3 gives the four potentials \(B_{(1)}^I\)
used below.\footnote{A duality frame specifies which four vector
potentials are treated as fundamental. The Lagrangian above and the
explicit solution and thermodynamics of Ref.~\cite{Ma:2026fourcharge}
use a mixed electric--magnetic frame, in which the four-charge seed
carries two electric and two magnetic charges. We present our solution
and electromagnetic charges in the electric frame based on
\(B_{(1)}^I\). The two frames are related by
\eqref{eq:seven-mixed-potentials}. Our general solution remains
dyonic in this frame.} The component Lagrangian and field equations are given in
Appendix~\ref{app:stu-eom}.

\subsection{The solution}
\label{sec:seven-solution}

We now present the rotating eight-charge solution, giving the metric,
gauge potentials and scalar fields in turn. At fixed gauging parameter
\(g>0\), the local solution has ten independent real parameters
\((m,a,\delta_1,\delta_2,\delta_3,\delta_4,\gamma,
                                      \vartheta_1,\vartheta_2,\vartheta_3)\).
Here \(m>0\) sets the mass scale and \(a>0\) controls rotation.
The four \(\delta_I\in\mathbb R\), \(I=1,\ldots,4\), are electric
charging boosts, while \(\gamma\in\mathbb R\) is an additional
magnetic charging boost. The three compact angles
\(\vartheta_i\in\mathbb R\), \(i=1,2,3\), act on the vector and
scalar fields through electromagnetic duality rotations and leave
the metric unchanged. These five boosts and three angles collectively
determine the four electric and four magnetic charges of
\eqref{eq:eight-fluxes}. Their local independence and the global
black-hole restrictions are discussed in
section~\ref{sec:eight-global}. All auxiliary coefficients below are
fixed by these ten parameters and introduce no additional independent
parameters.

We use Carter coordinates \((t_0,r,\sigma,u)\).\footnote{The subscript ``\(0\)''
distinguishes the Carter time \(t_0\) from the normalized time \(t\)
used for the horizon and thermodynamic quantities. They are related by
\(t=\sqrt{\alpha_+}\,t_0\), as in
\eqref{eq:eight-angular-identification}.} To express the metric
and matter fields compactly, we introduce the shorthand notation
\begin{equation}
\begin{aligned}
 s_I&=\sinh\delta_I\,, & c_I&=\cosh\delta_I\,,\\
 s_\gamma&=\sinh\gamma\,, & c_\gamma&=\cosh\gamma\,,\\
 c_{I_1\cdots I_k}&=\prod_{p=1}^kc_{I_p}\,, & s_{I_1\cdots I_k}&=\prod_{p=1}^ks_{I_p}\,,\\
 \mathfrak s_k&=\sum_{1\leq I_1<\cdots<I_k\leq4}
                        s_{I_1}^2\cdots s_{I_k}^2\,.
\end{aligned}
\label{eq:eight-parameters}
\end{equation}
Here \(1\leq k\leq4\), and the product indices are distinct.
The parameter-dependent coefficients are written in the notation
of Ref.~\cite{Chow:2014ungauged}:
\begin{equation}
\begin{aligned}
 w_4&=c_{123}s_4-s_{123}c_4\,,\\
 d_0&=c_\gamma s_{1234}\,,\\
 \mu_1&=1+\tfrac12\mathfrak s_1
       +\tfrac12s_\gamma^2(1+\mathfrak s_1-2s_4^2)\,,\\
 \mu_2&=s_\gamma c_4s_4\,,\\
 \nu_1&=-s_\gamma c_\gamma w_4\,,\\
 \nu_2&=c_\gamma(c_{1234}-s_{1234})\,,\\
 c_0&=s_\gamma^2s_4^2-c_\gamma^2
       (\mathfrak s_3+2\mathfrak s_4-2c_{1234}s_{1234})
       -s_\gamma^2c_\gamma^2w_4^2\,.
\end{aligned}
\label{eq:eight-charging-coefficients}
\end{equation}
On the patch \(\nu_2\ne0\), the remaining coefficients are
\begin{equation}
\begin{aligned}
 n&=-\frac{m\nu_1}{\nu_2}\,,
 &\mathfrak m&=m\mu_1+n\mu_2\,,\\
 V_1&=2(n\mu_1-m\mu_2)\,,&V_0&=2(m^2+n^2)c_0\,,\\
 L_1&=2(m\nu_2-n\nu_1)\,,&L_0&=4(m^2+n^2)d_0\,,
\end{aligned}
\label{eq:eight-radial-coefficients}
\end{equation}
where \(\mathfrak m\) is a
charge-dependent metric coefficient.\footnote{Here \(n\) is fixed by \(m\) and the charging parameters through
\eqref{eq:eight-radial-coefficients}. Namely, it is not an independent NUT
parameter of the type listed in table~\ref{tab:ads4-solutions}.} The metric functions are
\begin{equation}
\begin{aligned}
 R_{\rm s}&=r^2-2mr+a^2-n^2\,,
 &U_{\rm s}&=a^2-(u-n)^2\,,\qquad \mathscr D=R_{\rm s}-U_{\rm s}\,,\\
 K&=r^2+(4\mathfrak m-2m)r+2V_0\,,
 &L&=L_1r+L_0\,,\\
 J_{\rm a}&=u^2+(2V_1-2n)u\,,
 &W&=\sqrt{\mathscr D(K+J_{\rm a})+L^2}\,,\\
 R_g&=R_{\rm s}(1+g^2K)+g^2L^2\,,
 &U_g&=U_{\rm s}(1-g^2J_{\rm a})\,,\\
 E&=(1-g^2J_{\rm a})R_g-(1+g^2K)U_g\,,
 &O&=LU_g\,,\\
 \mathcal H&=(K+J_{\rm a})R_{\rm s}+L^2\,,
 &\mathcal T&=U_{\rm s}\mathcal H\,.
\end{aligned}
\label{eq:eight-metric-functions}
\end{equation}
The line element is
\begin{equation}
 \mathrm{d}s_4^2=-\frac{E}{W}\,\mathrm{d}t_0^2
 -\frac{2O}{W}\,\mathrm{d}t_0\mathrm{d}\sigma
 +\frac{\mathcal T}{W}\,\mathrm{d}\sigma^2
 +W\left(\frac{\mathrm{d}r^2}{R_g}
                    +\frac{\mathrm{d}u^2}{U_g}\right)\,.
\label{eq:eight-metric}
\end{equation}
The identity \(E\mathcal T+O^2=W^2R_gU_g\)
implies \(\det(g_{\mu\nu})=-W^2\) in Carter coordinates.
We take the positive branch of \(W\) in
\eqref{eq:eight-metric-functions} and use the spacetime orientation
fixed in section~\ref{sec:seven-theory}.

The three compact angles enter through the matrices
\begin{equation}
\begin{aligned}
 \mathsf H&=\frac12
 \begin{pmatrix}
 -1&1&-1&1\\1&1&1&1\\-1&-1&1&1\\1&-1&-1&1
 \end{pmatrix}\,,\\
 (\psi_1\,,\psi_2)&=
 (-\vartheta_1-\vartheta_2+\vartheta_3\,,
  -\vartheta_1+\vartheta_2-\vartheta_3)\,,\\
 (\psi_3\,,\psi_4)&=
 (\vartheta_1-\vartheta_2-\vartheta_3\,,
  \vartheta_1+\vartheta_2+\vartheta_3)\,,\\
 \mathsf U&=\mathsf H^T
      \operatorname{diag}(\cos\psi_1\,,\ldots\,,\cos\psi_4)\mathsf H\,,\\
 \mathsf V&=\mathsf H^T
      \operatorname{diag}(\sin\psi_1\,,\ldots\,,\sin\psi_4)\mathsf H\,.
\end{aligned}
\label{eq:seven-duality-matrices}
\end{equation}
With the coefficients and auxiliary functions specified in
Appendix~\ref{app:eight-auxiliary}, the four gauge potentials and their
four local Maxwell duals are
\begin{equation}
\begin{aligned}
 \begin{pmatrix}B_{(1)}^I\\C_{(1)I}\end{pmatrix}
 =\sum_{J=1}^4\Bigg\{&
 \begin{pmatrix}\mathsf U_{IJ}\\-\mathsf V_{IJ}\end{pmatrix}
 \Bigg[\left((1-g^2J_{\rm a})
       \frac{\mathscr D X_J+LY_J}{W^2}+g^2u v_{J1}\right)
                                      \mathrm dt_0\\
 &\qquad+\left(\mathfrak p^J(u-n)
       +\frac{U_{\rm s}[LX_J-(K+J_{\rm a})Y_J]}{W^2}\right)
                                      \mathrm d\sigma\Bigg]\\
 -{}&\begin{pmatrix}\mathsf V_{IJ}\\\mathsf U_{IJ}\end{pmatrix}
 \Bigg[\left((1-g^2J_{\rm a})
       \frac{\mathscr D\widetilde X_J+L\widetilde Y_J}{W^2}
                         -g^2u a_{J1}\right)\mathrm dt_0\\
 &\qquad+\left(-\mathfrak q_J(u-n)
       +\frac{U_{\rm s}[L\widetilde X_J-(K+J_{\rm a})\widetilde Y_J]}{W^2}
                           \right)\mathrm d\sigma\Bigg]\Bigg\}\,.
\end{aligned}
\label{eq:seven-potentials}
\end{equation}
Here the electric-frame field strengths and their Maxwell duals are defined as
\begin{equation}
 F^I:=\mathrm dB_{(1)}^I\,,\qquad
 G_I:=\mathrm dC_{(1)I}\,,\qquad I=1\,,\ldots\,,4\,.
\label{eq:eight-field-strength-definitions}
\end{equation}
The electric--magnetic constitutive relation and the correspondence
with the mixed frame of \eqref{eq:fourcharge-lagrangian} are given in
Appendix~\ref{app:stu-eom}. Magnetic potentials are understood on
overlapping angular patches.\footnote{For nonzero magnetic charge, regular north and south
representatives of the gauge potential differ by a gauge transformation
on the overlap. The dual potentials require analogous patching for
nonzero electric charge. The field strengths and charge integrals are
independent of the patch representatives.}

The six real scalar profiles are expressed in terms of the functions
\(N_i\) and \(\eta_i\) specified in
Appendix~\ref{app:eight-auxiliary}. On the patch \(N_i>0\), they are
\begin{equation}
\begin{aligned}
 e^{-\varphi_i}
 &=\frac{WN_i}
 {(N_i\cos\vartheta_i-\eta_i\sin\vartheta_i)^2
                                      +W^2\sin^2\vartheta_i}\,,\\
 \chi_i
 &=\frac{\eta_iN_i\cos(2\vartheta_i)
       +\tfrac12(N_i^2-\eta_i^2-W^2)\sin(2\vartheta_i)}
 {(N_i\cos\vartheta_i-\eta_i\sin\vartheta_i)^2
                                      +W^2\sin^2\vartheta_i}\,.
\end{aligned}
\label{eq:eight-scalars}
\end{equation}
Equations~\eqref{eq:eight-metric}, \eqref{eq:seven-potentials}
and \eqref{eq:eight-scalars}, with the auxiliary functions in
Appendix~\ref{app:eight-auxiliary},
specify the complete local field configuration. The angles
\(\vartheta_i\) do not enter the metric. Global black-hole
conditions impose the additional restrictions described below.

\subsection{Black-hole branches, charges and boundary data}
\label{sec:eight-global}

We next examine the horizons and axes of the local solution, before
evaluating its electromagnetic charges and boundary data. On a
non-extremal branch, the event horizon is located at \(r=r_+\), where
\(r_+\) is the largest simple zero of \(R_g\) and
\(R_g'(r_+)>0\).\footnote{Primes on radial functions denote
derivatives with respect to \(r\) at fixed solution parameters.}
A regular exterior requires \(W>0\), \(N_i>0\),
\(R_g>0\) and \(\mathcal T>0\) away from the axes, while the horizon
must satisfy \(1+g^2K(r_+)>0\) and \(L(r_+)\ne0\).
Sufficient conditions for exterior causality are given in
section~\ref{sec:eight-ctcs}. The double-root case will be treated in
section~\ref{sec:eight-extremal}.

Take \(u\in[u_-,u_+]\), where \(u_\pm=n\pm a\), with
\(1-g^2J_{\rm a}>0\) throughout the interval. The north axis is
regular for
\begin{equation}
 \alpha_\pm=1-g^2J_{\rm a}(u_\pm)\,,\qquad
 \Delta\sigma=\frac{2\pi}{a\sqrt{\alpha_+}}\,,\qquad
 t=\sqrt{\alpha_+}\,t_0\,,\qquad
 \phi=a\sqrt{\alpha_+}\,\sigma\,.
\label{eq:eight-angular-identification}
\end{equation}
so that \(\phi\) has period \(2\pi\). We set
\(G_{N}^{(4)}=1\). Let \(\mathcal S_r\) be the angular cut
at fixed \(t_0\) and radius \(r\), with
\(u\in[u_-,u_+]\), \(\sigma\sim\sigma+\Delta\sigma\), and
orientation \(\mathrm du\wedge\mathrm d\sigma\). The magnetic and
electric charges are defined by the asymptotic flux integrals
\begin{equation}
 P^I:=\frac{1}{16\pi}\lim_{r\to\infty}\int_{\mathcal S_r}F^I\,,
 \qquad
 Q_I:=\frac{1}{16\pi}\lim_{r\to\infty}\int_{\mathcal S_r}G_I\,,
 \qquad I=1\,,\ldots\,,4\,.
\label{eq:eight-flux-definitions}
\end{equation}
Evaluating these integrals using
\eqref{eq:seven-potentials} gives\footnote{In the charge relation below, bold symbols
denote four-component columns ordered by \(I=1,\ldots,4\).
The same convention applies to the field strengths and potentials in
\eqref{eq:eight-field-strength-definitions}, with any other indices
held fixed. The one-form components of the potentials are
denoted by \(B^I_\mu\) and \(C_{\mu I}\).}

\begin{equation}
 \begin{pmatrix}\boldsymbol P\\\boldsymbol Q_{\rm el}\end{pmatrix}
 =\frac{1}{4\sqrt{\alpha_+}}
  \begin{pmatrix}\mathsf U&\mathsf V\\-\mathsf V&\mathsf U\end{pmatrix}
  \begin{pmatrix}\boldsymbol{\mathfrak p}\\\boldsymbol{\mathfrak q}\end{pmatrix}\,.
\label{eq:eight-fluxes}
\end{equation}
At fixed \(m,a,g\), the map from
\((\delta_I,\gamma,\vartheta_i)\) to these charges has rank eight
on a generic patch, allowing the boundary sources and axial defect
data to vary.
A regular non-extremal branch with rank-eight physical
fluxes at fixed \(g,k_+,k_-\) is described in
Appendix~\ref{app:eight-fixed-spindle-family}.
This distinguishes a nonzero axial imbalance from variations of
the polar opening angles: the rank-eight branch need not vary its
cone angles.

A representative of the conformal boundary metric is
\begin{equation}
 h_\partial=-g^2(1-g^2J_{\rm a})\,\mathrm{d}t_0^2
 +\frac{\mathrm{d}u^2}{U_{\rm s}(1-g^2J_{\rm a})}
 +U_{\rm s}\,\mathrm{d}\sigma^2\,.
\label{eq:eight-boundary-metric}
\end{equation}
The boundary conformal geometry also depends on the axial imbalance.
For \(h_\partial\), let \(\nabla^\partial\),
\(\mathcal R^\partial_{ab}\) and \(\mathcal R_\partial\) denote its
Levi-Civita derivative, Ricci tensor and Ricci scalar, respectively,
with boundary indices \(a,b,c\) running over \((t_0,u,\sigma)\).
The Cotton tensor is
\begin{equation}
 C^\partial_{abc}:=
 \nabla^\partial_c\left(\mathcal R^\partial_{ab}
                   -\frac14\mathcal R_\partial h_{\partial ab}\right)
 -\nabla^\partial_b\left(\mathcal R^\partial_{ac}
                   -\frac14\mathcal R_\partial h_{\partial ac}\right)\,.
\label{eq:eight-boundary-cotton}
\end{equation}
Evaluation on \eqref{eq:eight-boundary-metric}, with indices raised
using \(h_\partial\), gives
\begin{equation}
 C^\partial_{abc}C_\partial^{abc}
       =9g^4V_1^2U_{\rm s}(1-g^2J_{\rm a})\,.
\label{eq:eight-boundary-cotton-norm}
\end{equation}
All components vanish when \(V_1=0\). In the physical angular
interior, \(U_{\rm s}>0\) and \(1-g^2J_{\rm a}>0\), so the norm
is strictly positive for \(V_1\ne0\). Thus the boundary metric is
locally conformally flat precisely on the axial balance locus.

On the balanced locus, \(\alpha_+=\alpha_-=1+g^2(n^2-a^2)>0\).
With two regular axes, the identification
\eqref{eq:eight-angular-identification}, and noncompact time, the
global round representative can be exhibited explicitly. Introduce
the boundary coordinates
\begin{equation}
 \cos\Theta=
 \frac{\sqrt{\alpha_+}(u-n)}{a\sqrt{1-g^2J_{\rm a}}}\,,
 \qquad
 \tau_\partial=g\sqrt{1+g^2n^2}\,t \,.
\label{eq:eight-boundary-round-coordinates}
\end{equation}
The positive, smooth Weyl rescaling gives the Einstein static universe
\begin{equation}
 \widehat h_\partial
 =\frac{(1+g^2n^2)\alpha_+}{1-g^2J_{\rm a}}\,h_\partial
 =-\mathrm d\tau_\partial^2+\mathrm d\Theta^2
                          +\sin^2\Theta\,\mathrm d\phi^2 \,.
\label{eq:eight-boundary-round-representative}
\end{equation}
Here \(u_+\) and \(u_-\) map to \(\Theta=0\) and \(\pi\),
respectively, where the \(2\pi\)-periodic \(\phi\) circles collapse
smoothly. Since the quotient does not identify time, this representative
is globally \(\mathbb R\times S^2\).
Generic eight-charge members therefore carry both an axial defect
and a non-conformally-flat boundary geometry. This geometric
condition is separate from the electromagnetic and scalar boundary
conditions.

For the potentials in \eqref{eq:seven-potentials}, the temporal
boundary values are

\begin{equation}
 \left.\begin{pmatrix}\boldsymbol B_{t_0}\\
                         \boldsymbol C_{t_0}\end{pmatrix}
                                                      \right|_{r\to\infty}
 =g^2u\begin{pmatrix}\mathsf U&\mathsf V\\-\mathsf V&\mathsf U\end{pmatrix}
       \begin{pmatrix}\boldsymbol v_1\\\boldsymbol a_1\end{pmatrix}\,.
\label{eq:eight-boundary-sources}
\end{equation}
Which combinations are held fixed as sources depends on the
electromagnetic boundary polarization.\footnote{This specifies which electromagnetic variables are fixed
in the variational problem: boundary gauge potentials in a Dirichlet
prescription, or conjugate electric data or mixed combinations in other
prescriptions, with suitable boundary terms~\cite{Marolf:2006boundary}.
It is distinct from the choice of bulk duality frame.}
\paragraph{Charge independence and axial balance.}

The south-axis conical factor is \(\sqrt{\alpha_-/\alpha_+}\) with $\alpha_--\alpha_+=4ag^2V_1$.
Both axes are smooth precisely when \(V_1=0\). On this balanced
locus, the charges obey
\begin{equation}
 \operatorname{Im}\prod_{I=1}^4
       [\mathsf H(\boldsymbol Q_{\rm el}+\mathrm{i}\boldsymbol P)]_I=0\,.
\label{eq:eight-balanced-charge-relation}
\end{equation}
This nontrivial quartic relation restricts the eight electromagnetic
charges to a charge locus of dimension at most seven. Consequently,
within the present family, two smooth axes are incompatible with
eight locally independent classical electromagnetic charges. The
restriction holds on all balanced branches, including those with
\(\gamma\ne0\), and persists when the mass and rotation parameters
or the common azimuthal period are varied. It does not constitute an obstruction
for black-hole families beyond this construction.

Generic eight-charge members retain the boundary sources and axial
defect. Their fields solve the gauged equations on regular patches
away from the defect core, subject to the exterior conditions above.
The \(\gamma=0\) subfamily and its boundary data are described in
section~\ref{sec:seven-subfamily}. Extremal and non-extremal
spindle examples with specified polar opening angles are collected
in Appendix~\ref{app:eight-spindle-example}.

\subsection{The \(\gamma=0\) seven-charge subfamily}
\label{sec:seven-subfamily}

We now consider the \(\gamma=0\) limit of the eight-charge solution.
In this limit, \(n=V_1=0\), so the axial balance condition is satisfied
and the solution reduces to a smooth-axis seven-charge subfamily.
The limiting fields are conveniently expressed using the notation
\begin{equation}
 \Xi=1-a^2g^2>0\,,\quad
 r_I=r+2ms_I^2\,,\quad
 r_{I_1\cdots I_k}=\prod_{p=1}^kr_{I_p}\,,\quad
 \Delta=r^2-2mr+a^2\,.
\label{eq:seven-parameters}
\end{equation}
The angular coefficients satisfy \(\alpha_+=\alpha_-=\Xi\), so the
identification \eqref{eq:eight-angular-identification} becomes
\begin{equation}
 u=a\cos\theta\,,\qquad t=\sqrt\Xi\,t_0\,,\qquad
 \phi=a\sqrt\Xi\,\sigma\,,\qquad \Delta\phi=2\pi\,,
\label{eq:eight-seven-limit}
\end{equation}
and both axes are smooth. Setting \(\gamma=0\) in the metric
\eqref{eq:eight-metric}, vector potentials \eqref{eq:seven-potentials}
and scalars \eqref{eq:eight-scalars} gives the complete field
configuration of this seven-charge subfamily. The four boosts
\(\delta_I\) and three compact angles \(\vartheta_i\) remain subject
to the horizon and exterior regularity conditions stated above.

The compact rotations leave the metric unchanged while generating
dyonic gauge fields and transforming the scalar profiles. The geometry
therefore coincides with that of the four-charge seed, and setting all
three angles to zero recovers its full field configuration
\cite{Ma:2026fourcharge,Wu:2026fourcharge}.

Smooth axes do not, however, imply source-free electromagnetic
boundary data. For general \(\vartheta_i\), the boundary pullback of
the electric-frame field strengths is
\begin{equation}
 \left.\boldsymbol{F}\right|_{\partial\,,\,\gamma=0}
 =\mathsf V\left(g^2\boldsymbol a_1\,\mathrm du\wedge\mathrm dt_0
             +\boldsymbol{\mathfrak q}\,\mathrm du\wedge\mathrm d\sigma\right)\,,
\label{eq:seven-boundary-gauge-curvature}
\end{equation}
where the coefficients on the right-hand side are evaluated at
\(\gamma=0\). This pullback is generically nonzero, so the smooth-axis
seven-charge family does not in general satisfy source-free Dirichlet
gauge boundary conditions in this electric frame.

The detailed limiting fields and the thermodynamic, BPS and causal
results are collected in Appendix~\ref{app:seven-subfamily}, together
with the supersymmetric entropy function. The following section studies
the general eight-charge horizon quantities, extremal geometry and
causal properties, using this subfamily for comparison.

\section{Properties of the eight-charge solution}
\label{sec:eight-properties}

We analyze the horizon quantities, extremal near-horizon geometry
and causal properties of the eight-charge solution with the angular
quotient and boundary data specified in
section~\ref{sec:eight-global}.

\subsection{Horizon quantities}
\label{sec:eight-horizon}

We first evaluate the entropy, temperature, angular velocity and
electromagnetic potentials at the outer horizon of the non-extremal
solution. On the branch specified in section~\ref{sec:eight-global},
it is convenient to use the horizon coefficients
\begin{equation}
 \beta_+=1+g^2K(r_+)>0\,,\qquad L_+=L(r_+)\ne0\,.
\label{eq:eight-horizon-coefficients}
\end{equation}
With the normalization \eqref{eq:eight-angular-identification}, and in
the conventions of section~\ref{sec:eight-global}, the entropy,
temperature and angular velocity are
\begin{equation}
 S_+=\frac{\pi|L_+|}{\sqrt{\alpha_+\beta_+}}\,,\qquad
 T_+=\frac{R_g'(r_+)\sqrt{\beta_+}}
                 {4\pi|L_+|\sqrt{\alpha_+}}\,,\qquad
 \Omega_+=\frac{a\beta_+}{L_+}\,.
\label{eq:eight-horizon-data}
\end{equation}
The charges are given in \eqref{eq:eight-fluxes}. For the normalized
horizon generator \(\xi_+:=\partial_t+\Omega_+\partial_\phi\),
we use the electric and magnetic horizon contractions
\[
 \Phi_{I+}:=\left.\xi_+^\mu B^I_\mu\right|_{r_+},\qquad
 \Psi_{I+}:=-\left.\xi_+^\mu C_{\mu I}\right|_{r_+}.
\]
For the potentials in \eqref{eq:seven-potentials}, these contractions
are constant on the horizon and evaluate to
\begin{equation}
\begin{aligned}
 \begin{pmatrix}\Phi_{I+}\\\Psi_{I+}\end{pmatrix}
 =&-\frac1{\sqrt{\alpha_+}}\sum_{J=1}^4\Bigg\{
 \begin{pmatrix}\mathsf U_{IJ}\\\mathsf V_{IJ}\end{pmatrix}
 \left[g^2(\mathfrak q_Jr_++v_{J0})
 +\frac{\beta_+}{L_+}(a_{J1}r_++a_{J0}+n\mathfrak p^J)\right]\\
 &+{}\begin{pmatrix}-\mathsf V_{IJ}\\\mathsf U_{IJ}\end{pmatrix}
 \Big[g^2\bigl(\mathfrak p^Jr_++2(m^2+n^2)\widetilde c_J\bigr)+\frac{\beta_+}{L_+}
 \bigl(v_{J1}r_++4(m^2+n^2)\widetilde d_J-n\mathfrak q_J\bigr)
 \Big]\Bigg\}\,.
\end{aligned}
\label{eq:eight-horizon-potentials}
\end{equation}

These constants refer to the stationary Carter representatives in
\eqref{eq:seven-potentials}. In the normalized azimuth, their polar
values are \(B_\phi^I|_{u_\pm}=\pm4P^I\) and
\(C_{\phi I}|_{u_\pm}=\pm4Q_I\), consistent with the orientation
and flux normalization in \eqref{eq:eight-flux-definitions}.
The regular north/south representatives are therefore
\begin{equation}
 B_{(1)}^{I\,,N/S}=B_{(1)}^I\mp4P^I\,\mathrm d\phi\,,
 \qquad
 C_{(1)I}^{N/S}=C_{(1)I}\mp4Q_I\,\mathrm d\phi\,,
\label{eq:eight-polar-gauge-patches}
\end{equation}
The upper/lower signs refer to north/south. On the overlap,
\(B_{(1)}^{I,N}-B_{(1)}^{I,S}=\mathrm d(-8P^I\phi)\),
and likewise for \(C_{(1)I}\) with \(P^I\) replaced by \(Q_I\).
For the lift of \(\xi_+\), choose the constant gauge parameters
\begin{equation}
 \lambda_{+\,,N/S}^I=-\Phi_{I+}\pm4\Omega_+P^I\,,
 \qquad
 \widetilde\lambda_{+\,,I}^{N/S}=\Psi_{I+}\pm4\Omega_+Q_I \,.
\label{eq:eight-horizon-gauge-lift}
\end{equation}
They give
\(\left.(\iota_{\xi_+}B_{(1)}^{I,N/S}
                  +\lambda_{+,N/S}^I)\right|_{r_+}=0\)
and the analogous zero for \(C_{(1)I}^{N/S}\) and
\(\widetilde\lambda_{+,I}^{N/S}\).
The overlap rule
\(\lambda_{+,N}^I-\lambda_{+,S}^I
       =-\mathcal L_{\xi_+}(-8P^I\phi)\), and its dual analogue,
makes these compensated contractions patch independent. The
uncompensated contractions remain constant in each stationary patch,
but their values shift by the displayed \(\Omega_+\)-dependent constants.

For the generic eight-charge branch, interpreting these horizon
potentials as thermodynamic chemical potentials requires a specified
gauge and boundary reference. An integrable Hamiltonian mass and
angular momentum, together with a first law incorporating variations
of the boundary sources and axial defect data, remain to be established
with a compatible boundary and defect
prescription~\cite{Iyer:1994noether,Appels:2017conical}.
This prescription must specify the electromagnetic polarization,
scalar boundary conditions and defect action, and account for
gauge-patch contributions.

The seven-charge quantities below use the normalization
of the four-charge seed of Ref.~\cite{Ma:2026fourcharge}.
Their first-law relation is a differential identity on
the solution parameter space. Its Hamiltonian interpretation requires
a compatible boundary prescription, as discussed in
Appendix~\ref{sec:seven-thermodynamics}. At
\(\gamma=0\), \(n=V_1=0\) and \(\alpha_+=\Xi\), the horizon
quantities and electromagnetic potentials reduce to those in
\eqref{eq:seven-horizon-thermodynamics}--\eqref{eq:seven-charges-potentials}.

The seed-normalized quantities in
\eqref{eq:seven-mass-spin} can equivalently be written as
\begin{equation}
 M=\frac{\mathfrak m}{\Xi}\,,\qquad
 J=\frac{aL_1}{2\Xi}\,.
\label{eq:seven-comparison-mass-spin}
\end{equation}
At fixed \(g\), the first law \eqref{eq:seven-first-law} reduces to
\begin{equation}
 \mathrm dM=T_+\,\mathrm dS_++\Omega_+\,\mathrm dJ
 +\sum_{I=1}^4\left(\Phi_{I+}\,\mathrm dQ_I
                         +\Psi_{I+}\,\mathrm dP^I\right)\,.
\label{eq:seven-comparison-first-law}
\end{equation}
The full seven-charge thermodynamic quantities, including the
pressure and its conjugate volume when \(g\) varies, are given in
Appendix~\ref{sec:seven-thermodynamics}.

\subsection{Extremal near-horizon geometry}
\label{sec:eight-extremal}

We now take the extremal limit and determine the resulting near-horizon
metric and matter fields. An ordinary extremal horizon corresponds to
a double root \(r_{\rm e}\) of \(R_g\), with
\(R_g''(r_{\rm e})>0\). The double-root conditions give
\begin{equation}
\begin{aligned}
 a^2&=n^2+2mr_{\rm e}-r_{\rm e}^2
             -\frac{g^2L(r_{\rm e})^2}{1+g^2K(r_{\rm e})}\,,\\
 0&=r_{\rm e}-m
     +\frac{g^2L(r_{\rm e})L_1}{1+g^2K(r_{\rm e})}
     -\frac{g^4L(r_{\rm e})^2(r_{\rm e}+2\mathfrak m-m)}
                  {[1+g^2K(r_{\rm e})]^2}\,.
\end{aligned}
\label{eq:eight-extremal-conditions}
\end{equation}
The near-horizon limit of the metric and matter fields can be written
explicitly for arbitrary \(\gamma\). The horizon coefficients are
\begin{equation}
\begin{aligned}
 \beta_{\rm e}&=1+g^2K(r_{\rm e})>0\,,\,\quad 
 L_{\rm e}=L(r_{\rm e})\ne0\,,\quad
 W_{\rm e}(u)=W(r_{\rm e}\,,u)\,,\\
 v_{\rm e}=&\tfrac12R_g''(r_{\rm e})=\beta_{\rm e}
       +4g^2(r_{\rm e}-m)(r_{\rm e}+2\mathfrak m-m)
       +g^2L_1^2-\frac{g^4L_{\rm e}^2}{\beta_{\rm e}}>0\,.
\end{aligned}
\label{eq:eight-nh-coefficients}
\end{equation}
In co-rotating coordinates \((\tau,\rho,\varphi,u)\), take
\(\varepsilon\to0^+\) with
\begin{equation}
 r=r_{\rm e}+\varepsilon\rho\,,\qquad
 t_0=\frac{|L_{\rm e}|}{\varepsilon v_{\rm e}\sqrt{\beta_{\rm e}}}
          \,\tau\,,\qquad
 \sigma=\frac{\varphi}{a\sqrt{\alpha_+}}
                 +\frac{\beta_{\rm e}}{L_{\rm e}}t_0\,.
\label{eq:eight-nh-scaling}
\end{equation}
Here \(\rho>0\) is the near-horizon radial coordinate and
\(\varphi\) has period \(2\pi\), with the angular identification
\eqref{eq:eight-angular-identification}. The resulting metric is
\begin{equation}
 \mathrm ds_{\rm nh}^2=\frac{W_{\rm e}}{v_{\rm e}}
 \left(-\rho^2\mathrm d\tau^2+\frac{\mathrm d\rho^2}{\rho^2}\right)
 +\frac{W_{\rm e}}{U_g}\,\mathrm du^2+\frac{U_gL_{\rm e}^2}{a^2\alpha_+\beta_{\rm e}W_{\rm e}}
          (\mathrm d\varphi+k_{\rm e}\rho\mathrm d\tau)^2\,,
\label{eq:eight-nh-metric}
\end{equation}
where the fibration coefficient is
\begin{equation}
 k_{\rm e}=
 \frac{a\sqrt{\alpha_+}
             [\beta_{\rm e}L_1
                  -2g^2(r_{\rm e}+2\mathfrak m-m)L_{\rm e}]}
      {v_{\rm e}|L_{\rm e}|\sqrt{\beta_{\rm e}}}\,.
\label{eq:eight-nh-rotation}
\end{equation}
With the extremal restriction
\(\Phi_{I,\rm e}=\left.\Phi_{I+}\right|_{r_+=r_{\rm e}}\),
the gauge shift
\(B_{(1)}^I\mapsto B_{(1)}^I-\Phi_{I,\rm e}\,\mathrm dt\) sets the electric potential on the extremal horizon to zero.

This is the extremal restriction of the lift in
\eqref{eq:eight-horizon-gauge-lift}. In a regular polar patch the same
operation reads
\(B_{(1)}^{I,N/S}\mapsto B_{(1)}^{I,N/S}
 -(\Phi_{I,\rm e}\mp4\Omega_{\rm e}P^I)\,\mathrm dt\),
where \(\Omega_{\rm e}=\left.\Omega_+\right|_{r_+=r_{\rm e}}\).
Since \(\phi=\varphi+\Omega_{\rm e}t\), the resulting
near-horizon potentials obey
\(B_{(1),\rm nh}^{I,N/S}=B_{(1),\rm nh}^I
 \mp4P^I\,\mathrm d\varphi\). Hence they have the same field strength
and magnetic flux. The dual potentials use the corresponding
\(\widetilde\lambda_{+,I}^{N/S}\) choice.
The resulting near-horizon
gauge potentials and field strengths are
\begin{equation}
\begin{aligned}
 B_{(1)\,,\rm nh}^I={}&b_{\rm e}^I(u)
           (\mathrm d\varphi+k_{\rm e}\rho\mathrm d\tau)
                      +e_{\rm e}^I\rho\mathrm d\tau\,,\\
 b_{\rm e}^I(u)={}&\frac{B_\sigma^I(r_{\rm e}\,,u)}{a\sqrt{\alpha_+}}\,,\\
 F_{\rm nh}^I=\mathrm dB_{(1)\,,\rm nh}^I={}&b_{\rm e}^{I\prime}(u)\,\mathrm du\wedge
           (\mathrm d\varphi+k_{\rm e}\rho\mathrm d\tau)
       +(e_{\rm e}^I+k_{\rm e}b_{\rm e}^I)\,
                              \mathrm d\rho\wedge\mathrm d\tau\,.
\end{aligned}
\label{eq:eight-nh-fields}
\end{equation}
Here \(I=1,\ldots,4\), and the prime on \(b_{\rm e}^I\) denotes
differentiation with respect to \(u\). The constants \(e_{\rm e}^I\) are
\begin{equation}
\begin{aligned}
 e_{\rm e}^I=\sum_{J=1}^4\Bigg\{&
 -\frac{|L_{\rm e}|}{v_{\rm e}\sqrt{\beta_{\rm e}}}
 \left[g^2(\mathsf U_{IJ}\mathfrak q_J-\mathsf V_{IJ}\mathfrak p^J)
 +\frac{\beta_{\rm e}}{L_{\rm e}}
          (\mathsf U_{IJ}a_{J1}-\mathsf V_{IJ}v_{J1})\right]\\
 &+\frac{k_{\rm e}}{a\sqrt{\alpha_+}}
 \Big[\mathsf U_{IJ}(a_{J1}r_{\rm e}+a_{J0}+n\mathfrak p^J)-\mathsf V_{IJ}
 (v_{J1}r_{\rm e}+4(m^2+n^2)\widetilde d_J-n\mathfrak q_J)
 \Big]\Bigg\}\,.
\end{aligned}
\label{eq:eight-nh-electric-coefficients}
\end{equation}
The coefficients in \eqref{eq:eight-nh-fields} and
\eqref{eq:eight-nh-electric-coefficients} use the symmetric monopole
representative of \eqref{eq:seven-potentials}. Writing the regular
north/south potentials in the same fiber form gives
\[
 b_{\rm e}^{I,N/S}=b_{\rm e}^I\mp4P^I\,,\qquad
 e_{\rm e}^{I,N/S}=e_{\rm e}^I\pm4k_{\rm e}P^I\,.
\]
The upper/lower signs denote the north/south patches. Thus
\(e_{\rm e}^{I,N/S}+k_{\rm e}b_{\rm e}^{I,N/S}
 =e_{\rm e}^I+k_{\rm e}b_{\rm e}^I\), and the field strength is
unchanged.
The Maxwell duals follow from \eqref{eq:eight-constitutive}, with
the couplings and Hodge dual evaluated on the near-horizon fields.
The six real near-horizon scalar profiles are the horizon values of
\eqref{eq:eight-scalars}
\begin{equation}
 \varphi_{i\,,\rm e}(u)=\varphi_i(r_{\rm e}\,,u)\,,\qquad
 \chi_{i\,,\rm e}(u)=\chi_i(r_{\rm e}\,,u)\,,\qquad i=1\,,2\,,3\,.
\label{eq:eight-nh-scalars}
\end{equation}
We use \(N_{i,\rm e}(u)=N_i(r_{\rm e},u)\) in the regularity
conditions below.

For a regular near-horizon patch, \(\beta_{\rm e}>0\),
\(v_{\rm e}>0\), \(L_{\rm e}\ne0\), \(\alpha_\pm>0\),
\(W_{\rm e}>0\) and \(N_{i,\rm e}>0\) throughout
\(u\in[u_-,u_+]\), and \(U_g>0\) in the interior. At either pole,
the proper distance \(y_\perp\) from the axis gives
\begin{equation}
 \frac{W_{\rm e}}{U_g}\,\mathrm du^2=\mathrm dy_\perp^2\,,\qquad
 \frac{U_gL_{\rm e}^2}{a^2\alpha_+\beta_{\rm e}W_{\rm e}}
             =c_\pm^2y_\perp^2+O(y_\perp^4)\,,\qquad
 c_+=1\,,\quad c_-=\sqrt{\frac{\alpha_-}{\alpha_+}}\,.
\label{eq:eight-fixed-polar-angles}
\end{equation}
The scalar profiles are regular at the poles. The symmetric monopole
representative obeys \(b_{\rm e}^I(u_\pm)=\pm4P^I\), with gauge
potentials taken on regular overlapping patches. Both axes are smooth
when \(c_-=1\); otherwise the south pole retains the prescribed conical
defect. These local conditions must be supplemented by a regular
exterior as specified in section~\ref{sec:eight-global}.

The temperature vanishes, and the horizon area gives the extremal
entropy
\begin{equation}
 T_{\rm e}=0\,,\qquad
 S_{\rm e}=\frac{\pi|L_{\rm e}|}{\sqrt{\alpha_+\beta_{\rm e}}}\,.
\label{eq:eight-extremal-area}
\end{equation}
This agrees with \eqref{eq:eight-horizon-data} at \(r_+=r_{\rm e}\).
An explicit extremal spindle solution within the eight-charge family,
together with its angular quotient and entropy rescaling, is collected
in Appendix~\ref{app:eight-extremal-spindle-example}.

The construction above establishes extremality without assuming
supersymmetry. The additional BPS restrictions established here
apply to the \(\gamma=0\) seven-charge branch of
Appendix~\ref{sec:seven-limits}. In particular, the compact rotations
must preserve the fixed FI vector and satisfy
\cite{Cacciatori:2016duality}
\begin{equation}
 \psi_2=-\vartheta_1+\vartheta_2-\vartheta_3=0\,,
 \qquad \vartheta_2=\vartheta_1+\vartheta_3\,.
\label{eq:eight-fi-subgroup}
\end{equation}
This angle constraint selects FI-preserving rotations, which is not a
sufficient condition for supersymmetry at nonzero \(\gamma\).
Establishing a \(\gamma\ne0\) BPS branch requires solving the local
Killing-spinor equations and imposing the global spinor conditions.

\subsection{Closed timelike curves}
\label{sec:eight-ctcs}

We finally examine the conditions excluding closed timelike curves in
the exterior away from the defect core. The periodic azimuthal orbits
must be spacelike, requiring \(g_{\phi\phi}>0\) away from the axes.
On the real scalar patch \(W>0\), \(N_i>0\), with \(R_g,U_g>0\)
outside the horizon, this requirement is equivalent to
\(\mathcal T>0\), as follows from the metric components
\begin{equation}
 g_{\phi\phi}=\frac{\mathcal T}{a^2\alpha_+W}\,,\qquad
 g^{tt}=-\frac{\alpha_+\mathcal T}{WR_gU_g}\,.
\label{eq:eight-causal-components}
\end{equation}
In the angular range of section~\ref{sec:eight-global},
\(1+g^2K>0\) and \(K+J_{\rm a}>0\) throughout the exterior
are sufficient for \(\mathcal H,\mathcal T>0\), since
\begin{equation}
 \mathcal H
 =\frac{(K+J_{\rm a})R_g+L^2(1-g^2J_{\rm a})}{1+g^2K}\,.
\label{eq:eight-causal-identity}
\end{equation}
An alternative radial criterion follows from the derivative at fixed \(u\)
\begin{equation}
\begin{aligned}
 \partial_r\mathcal H={}&
 2(r+2\mathfrak m-m)R_{\rm s}
 +2(r-m)(K+J_{\rm a})+2LL_1\,.
\end{aligned}
\label{eq:eight-ctc-monotonicity}
\end{equation}
If \(\partial_r\mathcal H>0\) for all \(r\geq r_+\) and all
allowed \(u\), the positive horizon value gives
\begin{equation}
 \mathcal H(r\,,u)\geq\mathcal H(r_+\,,u)
 =\frac{L_+^2[1-g^2J_{\rm a}(u)]}{\beta_+}>0\,.
\label{eq:eight-ctc-horizon-bound}
\end{equation}
Under either sufficient criterion, \(g_{\phi\phi}>0\) on and
outside the horizon away from the axes, and \(g^{tt}<0\) outside.
Since the quotient does not identify \(t\), the exterior is free
of CTCs away from the defect core.

For comparison with the BPS branch, we next consider a distinguished
Killing vector. Writing \(u=n+a\cos\theta\), with
\(0<\theta<\pi\), the Killing vector
\(K_+:=\partial_t+g\partial_\phi\) has norm
\begin{equation}
\begin{aligned}
 K_+^2={}&-\frac{WR_g(1-g^2J_{\rm a})}{\alpha_+\mathcal H}
 +\frac{\mathcal H\sin^2\theta}{\alpha_+W}
 \left[\frac{a(1-g^2J_{\rm a})L}{\mathcal H}-g\right]^2\,,\\
 \left.K_+^2\right|_{r_+}
 ={}&\left.g_{\phi\phi}\right|_{r_+}(\Omega_+-g)^2\,.
\end{aligned}
\label{eq:eight-ctc-killing-norm}
\end{equation}
Thus the additional requirement that \(K_+\) be non-spacelike on
a horizon with \(g_{\phi\phi}>0\) gives \(\Omega_+=g\),
while the general CTC criteria above impose no such equality.
The supersymmetric interpretation is discussed in
section~\ref{sec:eight-extremal}, and the radial monotonicity and BPS
Killing-vector analysis in the seven-charge limit are given in
Appendix~\ref{sec:seven-ctcs}.

\section{Discussion}\label{sec:discussion}

We have constructed exact non-extremal rotating AdS\(_4\) solutions in
gauged STU supergravity carrying four electric and four magnetic
charges. 
At fixed gauge coupling, the local family contains ten parameters.
A regular non-extremal branch has eight locally independent
electromagnetic fluxes even at fixed spindle orders, while boundary
sources vary. We have determined the horizon quantities and extremal
near-horizon fields, and obtained sufficient conditions excluding
closed timelike curves in the exterior away from the defect core.
The \(\gamma=0\) limit gives the smooth-axis seven-charge family,
whose seed-normalized thermodynamic quantities obey a first-law
identity on the solution parameter space. On its
established BPS branch, the entropy functional in electric seed
variables reproduces the Bekenstein--Hawking entropy. 
For horizon variations preserving the boundary sources and defect
data, the regulated local central coefficients give a formal Cardy
area matching, conditional on a globally integrable horizon phase
space, a suitable quantum realization and the applicability of the
Cardy formula.

These solutions extend the gravitational backgrounds available for
studying charged rotating states in three-dimensional field theories
with nontrivial electromagnetic sources and axial defects. The freedom
to vary the charges is restricted by global regularity: within this
family, smoothness of both axes imposes a quartic charge relation,
leaving at most seven independent electromagnetic charges. The generic
eight-charge solution therefore retains an axial conical defect. Its
holographic interpretation must account for the boundary gauge
background and the associated defect data.

For the general eight-charge branch, further work is needed to
choose the boundary and defect ensemble, integrate the corresponding
covariant charges, and complete its thermodynamic description.
It is tempting to seek a complete rotating eight-charge black-hole
solution with smooth axes and a round
conformal boundary, subject to explicitly specified electromagnetic and
scalar boundary conditions admitting nonzero magnetic flux in a specified
electromagnetic polarization. The explicit local solution obtained here suggests that we
are close to this goal, although achieving it requires going beyond
the present family, satisfying the boundary conditions and establishing
eight independent physical charges.
A central difficulty is that, in the present family,
full rank-eight charge freedom is tied to axial imbalance and a nonzero
boundary Cotton tensor, while imposing axial balance enforces the quartic
charge constraint. An extension must avoid this constraint while satisfying
the chosen boundary conditions and maintaining a regular horizon and exterior.
It would also be interesting to include acceleration, following the gauged
STU C-metrics and rotating supersymmetric spindle
solutions~\cite{Lu:2014sza,Ferrero:2020twa,Ferrero:2021ovq}.

Finally, the microscopic interpretation of the established
\(\gamma=0\) BPS branch requires matching its supersymmetric boundary
background and electromagnetic polarization to the field-theory
ensemble, taking account of the boundary gauge curvature
\eqref{eq:seven-boundary-gauge-curvature}. The generalized
superconformal index of ABJM theory with
background flavor magnetic fluxes~\cite{Kapustin:2011jm} provides a
natural framework for this comparison, building on
Refs.~\cite{Choi:2019zpz,Nian:2019pxj,Hosseini:2019iad}.

\acknowledgments
We thank Jun Nian and Jie Ren for many helpful discussions and acknowledge the assistance of ChatGPT in the computations and the preparation of the manuscript draft.
Y.-T. Wang is grateful to Hong L\"u and Jun Nian for introducing the related topics and for their helpful suggestions and guidance.
W.Z. thanks Mirjam Cveti\v{c} and Hong L\"u for helpful discussions of the paper~\cite{Chong:2004na} at conferences over the years, and Hong L\"u for encouraging her to further explore supergravity.
She is grateful to Jie Ren for introducing her to STU supergravity and exact solutions of Einstein's equations.
She also thanks Jun Nian and Leopoldo A. Pando Zayas for collaborations on related projects, and Jun Nian for his continued encouragement to publish this work.
W.Z. thanks the organizer of the Lunch Seminar at the Yau Center of Southeast University for the opportunity to orally report this work in October.
Y.W.  is supported by the NSFC grants No. 12247103 and 12605125.

\appendix

\section{Lagrangian, field equations and auxiliary functions}
\label{app:stu-eom}

This appendix collects the component Lagrangian, field equations,
duality conventions and auxiliary functions used for the eight-charge
solution, in the conventions of section~\ref{sec:seven-theory}.

\subsection{Lagrangian, duality conventions and field equations}
\label{app:electric-equations}

The component Lagrangian is\footnote{This electric-frame expression
is obtained by dualizing the first and third vectors in the mixed-frame
Lagrangian \eqref{eq:fourcharge-lagrangian}, following the conventions
of Ref.~\cite{Chong:2004na}.}
\begin{align}
 \mathcal L_{\rm el}={}&\sqrt{-\det(g_{\mu\nu})}\biggl[
 R-\frac12\sum_{i=1}^3(\partial\varphi_i)^2
 -\frac12\sum_{i=1}^3e^{2\varphi_i}(\partial\chi_i)^2
 -\frac14\mathsf K_{IJ}\,
                   F^I_{\mu\nu}F^{J\mu\nu}-V\biggr]\nonumber\\
 &-\frac18\mathsf\Theta_{IJ}
       \widetilde\epsilon^{\mu\nu\rho\lambda}
       F^I_{\mu\nu}F^J_{\rho\lambda}\,.
\label{eq:appA-component-lagrangian}
\end{align}
The exterior-derivative definitions in
section~\ref{sec:seven-solution} give the component expressions
\[
\begin{aligned}
 F^I_{\mu\nu}
 &=\partial_\mu B^I_\nu-\partial_\nu B^I_\mu
   =2\partial_{[\mu}B^I_{\nu]},\\
 G_{I\mu\nu}
 &=\partial_\mu C_{\nu I}-\partial_\nu C_{\mu I}
   =2\partial_{[\mu}C_{\nu]I}.
\end{aligned}
\]
Here \(I,J=1,2,3,4\), and \(i=1,2,3\). Repeated vector indices
are summed, and each scalar equation holds separately for \(i\), with no
sum over \(i\).
The potential \(V\) is the one specified in
\eqref{eq:fourcharge-potential}.
The gauge couplings are specified by
\(\kappa_i=\chi_i e^{\varphi_i}\),
\(h_i=1+\kappa_i^2\), and
\(\mathcal Z=1+\kappa_1^2+\kappa_2^2+\kappa_3^2
                -2\mathrm{i}\kappa_1\kappa_2\kappa_3\).
In the ordering of \(B_{(1)}^I\),
\begin{equation}
\begin{aligned}
 \mathsf E&=\operatorname{diag}\left(
 e^{(\varphi_1-\varphi_2+\varphi_3)/2}\,,
 e^{(-\varphi_1+\varphi_2+\varphi_3)/2}\,,
 e^{(\varphi_1+\varphi_2-\varphi_3)/2}\,,
 e^{-(\varphi_1+\varphi_2+\varphi_3)/2}\right)\,,\\
 \mathsf\Theta+\mathrm{i}\mathsf K&=\frac{1}{\mathcal Z}\mathsf E
 \begin{pmatrix}
 \mathrm{i}h_2&-\kappa_3+\mathrm{i}\kappa_1\kappa_2
   &-\kappa_1+\mathrm{i}\kappa_2\kappa_3
   &h_2(-\kappa_2+\mathrm{i}\kappa_1\kappa_3)\\
 -\kappa_3+\mathrm{i}\kappa_1\kappa_2&\mathrm{i}h_1
   &\kappa_2-\mathrm{i}\kappa_1\kappa_3
   &h_1(\kappa_1-\mathrm{i}\kappa_2\kappa_3)\\
 -\kappa_1+\mathrm{i}\kappa_2\kappa_3
   &\kappa_2-\mathrm{i}\kappa_1\kappa_3&\mathrm{i}h_3
   &h_3(\kappa_3-\mathrm{i}\kappa_1\kappa_2)\\
 h_2(-\kappa_2+\mathrm{i}\kappa_1\kappa_3)
   &h_1(\kappa_1-\mathrm{i}\kappa_2\kappa_3)
   &h_3(\kappa_3-\mathrm{i}\kappa_1\kappa_2)
   &\mathrm{i}h_1h_2h_3
 \end{pmatrix}\mathsf E\,.
\end{aligned}
\label{eq:appA-electric-coupling}
\end{equation}

The field strengths introduced in section~\ref{sec:seven-solution}
obey
\begin{equation}
 G_I=\mathsf\Theta_{IJ}F^J
                         +\mathsf K_{IJ}\star F^J\,,
\label{eq:eight-constitutive}
\end{equation}
with gauge couplings given in \eqref{eq:appA-electric-coupling}.
In the mixed frame of
\eqref{eq:fourcharge-lagrangian},
\begin{equation}
 (A_{(1)1}\,,A_{(1)2}\,,\mathcal A_{(1)}^1\,,\mathcal A_{(1)}^2)
 =(-C_{(1)1}\,,B_{(1)}^2\,,-C_{(1)3}\,,B_{(1)}^4)\,.
\label{eq:seven-mixed-potentials}
\end{equation}
These dualization signs and the axions in \eqref{eq:eight-scalars} use the orientation
\(\epsilon_{t_0r\sigma u}=W\).

The equations of motion can be written as
\begin{small}
\begin{align}
 &R_{\mu\nu}-\frac12g_{\mu\nu}R
   =\frac12\left(T^A_{\mu\nu}+T^\phi_{\mu\nu}\right)\,,
\label{eq:appA-einstein}\\
 &\sqrt{-\det(g_{\mu\nu})}\,\nabla_\mu
       \left(\mathsf K_{IJ}F^{J\mu\nu}\right)
 +\frac12\widetilde\epsilon^{\mu\nu\rho\lambda}
       \partial_\mu
       \left(\mathsf\Theta_{IJ}F^J_{\rho\lambda}\right)=0\,,
\label{eq:appA-maxwell}\\
 &\frac1{\sqrt{-\det(g_{\mu\nu})}}\partial_\mu
       \left(\sqrt{-\det(g_{\mu\nu})}\,\partial^\mu\varphi_i\right)
 -e^{2\varphi_i}(\partial\chi_i)^2
 -\partial_{\varphi_i}V
 -\frac14(\partial_{\varphi_i}\mathsf K_{IJ})
                       F^I_{\rho\lambda}F^{J\rho\lambda}
 \nonumber\\[-2pt]
 &\hspace{15mm}
 -\frac1{8\sqrt{-\det(g_{\mu\nu})}}
       (\partial_{\varphi_i}\mathsf\Theta_{IJ})
       \widetilde\epsilon^{\mu\nu\rho\lambda}
       F^I_{\mu\nu}F^J_{\rho\lambda}=0\,,
\label{eq:appA-dilaton}\\
 &\frac1{\sqrt{-\det(g_{\mu\nu})}}\partial_\mu
       \left(e^{2\varphi_i}\sqrt{-\det(g_{\mu\nu})}\,\partial^\mu\chi_i\right)
 -\partial_{\chi_i}V
 -\frac14(\partial_{\chi_i}\mathsf K_{IJ})
                       F^I_{\rho\lambda}F^{J\rho\lambda}
 \nonumber\\[-2pt]
 &\hspace{15mm}
 -\frac1{8\sqrt{-\det(g_{\mu\nu})}}
       (\partial_{\chi_i}\mathsf\Theta_{IJ})
       \widetilde\epsilon^{\mu\nu\rho\lambda}
       F^I_{\mu\nu}F^J_{\rho\lambda}=0\,,
\label{eq:appA-axion}
\end{align}
\end{small}
where
\begin{align}
 T^A_{\mu\nu}={}&\mathsf K_{IJ}
       \left(F^I_{\mu\rho}
                         F^J_{\nu}{}^\rho
             -\frac14g_{\mu\nu}
                         F^I_{\rho\lambda}F^{J\rho\lambda}\right)\,,
\label{eq:appA-gauge-stress}\\
 T^\phi_{\mu\nu}={}&\sum_{i=1}^3\biggl[
       \partial_\mu\varphi_i\partial_\nu\varphi_i
       -\frac12g_{\mu\nu}(\partial\varphi_i)^2
       +e^{2\varphi_i}\left(\partial_\mu\chi_i\partial_\nu\chi_i
               -\frac12g_{\mu\nu}(\partial\chi_i)^2\right)\biggr]
       -g_{\mu\nu}V\,.
\label{eq:appA-scalar-stress}
\end{align}

\subsection{Auxiliary functions of the eight-charge solution}
\label{app:eight-auxiliary}

The following coefficients and functions complete the explicit field
configuration in section~\ref{sec:seven-solution}. The charging
parameters and radial coefficients are those of
\eqref{eq:eight-parameters}--\eqref{eq:eight-radial-coefficients}.

For the vector fields, \(\partial_I\) acts on \(\delta_I\) at
fixed \(m,n,\gamma\), before imposing
\(n=-m\nu_1/\nu_2\). The coefficients are
\begin{equation}
\begin{aligned}
 \mathfrak q_I&=2(m\partial_I\mu_1+n\partial_I\mu_2)\,,
 &\mathfrak p^I&=-2(m\partial_I\nu_1+n\partial_I\nu_2)\,,\\
 a_{I1}&=-2m\partial_I\nu_2+2n\partial_I\nu_1\,,
 &a_{I0}&=-4(m^2+n^2)\partial_Id_0\,,\\
 v_{I1}&=2(n\partial_I\mu_1-m\partial_I\mu_2)\,,
 &v_{I0}&=2(m^2+n^2)\partial_Ic_0\,.
\end{aligned}
\label{eq:eight-vector-coefficients}
\end{equation}
For \(I=1,2,3\) and
\(\{J,K\}=\{1,2,3\}\setminus\{I\}\), with no sum over these labels,
the dual coefficients are
\begin{equation}
\begin{aligned}
 \widetilde c_I&=2s_\gamma c_\gamma\bigl[
 c_\gamma^2s_I^2(s_Is_4c_Jc_K-c_Ic_4s_Js_K)
                 +s_\gamma^2s_Is_4c_Jc_K\bigr]\,,
 &\widetilde d_I&=0\,,\\
 \widetilde c_4&=2s_\gamma c_\gamma\bigl[
 c_\gamma^2s_4^2(s_{1234}-c_{1234})
                    +s_{1234}(2+s_\gamma^2)\bigr]\,,
 &\widetilde d_4&=s_\gamma s_4^2\,.
\end{aligned}
\label{eq:eight-dual-coefficients}
\end{equation}
The linear functions entering \eqref{eq:seven-potentials} are
\begin{equation}
\begin{aligned}
 X_I&=\mathfrak q_Ir+v_{I1}u+v_{I0}\,,
 &Y_I&=-a_{I1}r-a_{I0}-\mathfrak p^Iu\,,\\
 \widetilde X_I&=\mathfrak p^Ir-a_{I1}u+2(m^2+n^2)\widetilde c_I\,,
 &\widetilde Y_I&=\mathfrak q_Iu-v_{I1}r-4(m^2+n^2)\widetilde d_I\,.
\end{aligned}
\label{eq:eight-vector-functions}
\end{equation}

For \(i=1,2,3\) and
\(\{j,k\}=\{1,2,3\}\setminus\{i\}\), the scalar auxiliary functions are
\begin{equation}
\begin{aligned}
 \rho_i&=c_\gamma^2(s_j^2+s_k^2)+s_\gamma^2\,,\\
 \upsilon_i&=c_jc_kc_4s_i-s_js_ks_4c_i\,,\\
 \lambda_i&=c_ic_4s_js_k-c_jc_ks_is_4\,,\\
 \widehat N_i&=r^2+u^2+2\rho_i(mr+nu)
                      +4(m^2+n^2)c_\gamma^2s_j^2s_k^2\,,\\
 \widehat f_i&=2c_\gamma\bigl[\lambda_i(mu-nr)
           -s_\gamma\upsilon_i(mr+nu)
           -2(m^2+n^2)s_\gamma c_is_js_ks_4\bigr]\,,\\
 (N_1\,,N_2\,,N_3)&=(\widehat N_2\,,\widehat N_1\,,\widehat N_3)\,,\\
 (\eta_1\,,\eta_2\,,\eta_3)&=(-\widehat f_2\,,\widehat f_1\,,-\widehat f_3)\,.
\end{aligned}
\label{eq:eight-scalar-functions}
\end{equation}
These functions enter the scalar profiles \eqref{eq:eight-scalars}.

\section{Results for the seven-charge subfamily}
\label{app:seven-subfamily}

This appendix collects the results specific to the \(\gamma=0\)
seven-charge subfamily of the solution in section~\ref{sec:seven-charge}.
We describe the seven-charge limit and collect its thermodynamics,
BPS conditions, causal properties and supersymmetric entropy function.
Away from the BPS limit, the four boosts and three compact angles are
subject only to the horizon and exterior regularity conditions specified above.

\subsection{Metric functions and limiting fields}
\label{app:seven-metric-functions}

For the recovered subfamily, the new auxiliary quantities are
\begin{equation}
\begin{aligned}
 d_1&=c_{13}s_{24}-c_{24}s_{13}\,,\qquad
 \mathcal P=r_{1234}\,,\\
 \mathcal K&=r_{13}+r_{24}-4m^2d_1^2\,,\\
 \mathcal F&=\mathcal P+a^2\mathcal K+a^4\,,\qquad
 \Delta_\theta=1-g^2a^2\cos^2\theta\,.
\end{aligned}
\label{eq:seven-metric-functions}
\end{equation}
The remaining functions retain their definitions in
\eqref{eq:eight-metric-functions}. We now derive their restrictions to
\(\gamma=0\), using \(u=a\cos\theta\) and the quantities in
\eqref{eq:seven-parameters}. Substituting \(\gamma=0\) into
\eqref{eq:eight-charging-coefficients} and
\eqref{eq:eight-radial-coefficients} gives
\[
\begin{aligned}
 \left.\mathfrak m\right|_{\gamma=0}
 &=m\left(1+\tfrac12\mathfrak s_1\right),\\
 \left.V_0\right|_{\gamma=0}
 &=-2m^2(\mathfrak s_3+2\mathfrak s_4-2c_{1234}s_{1234}),\\
 \left.L_1\right|_{\gamma=0}&=2m(c_{1234}-s_{1234}),\qquad
 \left.L_0\right|_{\gamma=0}=4m^2s_{1234}.
\end{aligned}
\]
Consequently, the linear function \(L\) reduces as
\[
\begin{aligned}
 \left.L\right|_{\gamma=0}
 &=\left.(L_1r+L_0)\right|_{\gamma=0}\\
 &=2m(c_{1234}-s_{1234})r+4m^2s_{1234}\\
 &=2m\left[rc_{1234}-(r-2m)s_{1234}\right].
\end{aligned}
\]
To reduce the other functions, expand the products using
\(c_I^2=1+s_I^2\). This gives
\[
\begin{aligned}
 d_1^2
 &=c_{13}^2s_{24}^2+c_{24}^2s_{13}^2-2c_{1234}s_{1234}\\
 &=s_{13}^2+s_{24}^2+\mathfrak s_3+2\mathfrak s_4
                                      -2c_{1234}s_{1234},\\
 c_{1234}^2&=\prod_{I=1}^4(1+s_I^2)
                         =1+\sum_{k=1}^4\mathfrak s_k.
\end{aligned}
\]
Using the first identity and \(r_I=r+2ms_I^2\) from
\eqref{eq:seven-parameters}, the new polynomial \(\mathcal K\) obeys
\[
\begin{aligned}
 \mathcal K
 &=r_{13}+r_{24}-4m^2d_1^2\\
 &=2r^2+2m\mathfrak s_1r
   -4m^2(\mathfrak s_3+2\mathfrak s_4-2c_{1234}s_{1234})\\
 &=\left.K\right|_{\gamma=0}+r^2-2mr.
\end{aligned}
\]
The last equality follows by substituting the reduced
\(\mathfrak m\) and \(V_0\) into \eqref{eq:eight-metric-functions}.
Similarly, substituting the reduced \(K\) and \(L\), using the
second product identity and \(s_{1234}^2=\mathfrak s_4\), yields
\[
\begin{aligned}
 \left.[(r^2-2mr)K+L^2]\right|_{\gamma=0}
 &=r^4+2m\mathfrak s_1r^3+4m^2\mathfrak s_2r^2
                          +8m^3\mathfrak s_3r+16m^4\mathfrak s_4\\
 &=\prod_{I=1}^4(r+2ms_I^2)\\
 &=\mathcal P.
\end{aligned}
\]
In the following right-hand sides, \(K\) and \(L\) are evaluated at
\(\gamma=0\). Substituting the relation for \(\mathcal K\) into
\eqref{eq:seven-metric-functions} first gives
\[
\begin{aligned}
 \mathcal F
 &=\mathcal P+a^2\mathcal K+a^4\\
 &=\mathcal P+a^2(K+r^2-2mr)+a^4\\
 &=\mathcal P+a^2K+a^2\Delta.
\end{aligned}
\]
Using these identities, \(\Xi=1-a^2g^2\) and
\(\Delta=r^2-2mr+a^2\), the radial and angular functions in
\eqref{eq:eight-metric-functions} reduce as
\[
\begin{aligned}
 \left.R_g\right|_{\gamma=0}
 &=\Delta(1+g^2K)+g^2L^2\\
 &=\Xi\Delta+g^2[\Delta(K+a^2)+L^2]\\
 &=\Xi\Delta+g^2(\mathcal P+a^2\mathcal K+a^4)\\
 &=\Xi\Delta+g^2\mathcal F,\\[4pt]
 \left.\mathcal H\right|_{\gamma=0,\,u=a\cos\theta}
 &=(K+a^2\cos^2\theta)\Delta+L^2\\
 &=\mathcal P+a^2K+a^2\cos^2\theta\,\Delta\\
 &=\mathcal F-a^2\sin^2\theta\,\Delta.
\end{aligned}
\]
Finally, \(U_{\rm s}=a^2-u^2\), \(J_{\rm a}=u^2\) and
\(\mathscr D=r^2-2mr+u^2\) on this subfamily, so
\[
\begin{aligned}
 \left.U_g\right|_{\gamma=0,\,u=a\cos\theta}
 &=(a^2-a^2\cos^2\theta)(1-g^2a^2\cos^2\theta)\\
 &=a^2\sin^2\theta\,\Delta_\theta,\\[4pt]
 \left.W^2\right|_{\gamma=0,\,u=a\cos\theta}
 &=(r^2-2mr+a^2\cos^2\theta)(K+a^2\cos^2\theta)+L^2\\
 &=\mathcal P+a^2\cos^2\theta(K+r^2-2mr)+a^4\cos^4\theta\\
 &=\mathcal P+a^2\cos^2\theta\,\mathcal K+a^4\cos^4\theta.
\end{aligned}
\]
Together, these restrictions reduce \eqref{eq:eight-metric} to
\begin{equation}
 \mathrm{d}s_4^2=
 -\frac{WR_g\Delta_\theta}{\Xi\mathcal H}\,\mathrm{d}t^2
 +\frac{\mathcal H\sin^2\theta}{\Xi W}
       \left(\mathrm{d}\phi
             -\frac{a\Delta_\theta L}{\mathcal H}
                       \,\mathrm{d}t\right)^2
 +W\left(\frac{\mathrm{d}r^2}{R_g}
                +\frac{\mathrm{d}\theta^2}{\Delta_\theta}\right)\,.
\label{eq:seven-metric}
\end{equation}
All eight potentials and all six real scalar fields reduce to those
of the \(\gamma=0\) seven-charge subfamily described in
section~\ref{sec:seven-subfamily}. With the angular identification
\eqref{eq:eight-seven-limit}, both axes are smooth. Setting the three
angles to zero further gives the
four-charge solution of
Refs.~\cite{Ma:2026fourcharge,Wu:2026fourcharge}.\footnote{In the mixed
frame of Ref.~\cite{Ma:2026fourcharge}, \(A_{(1)1}\) and
\(\mathcal A_{(1)}^1\) carry magnetic charges
\(P_{\rm mix}^1,P_{\rm mix}^3\), respectively.
Dualizing this pair gives the four-electric-charge frame used here.}
In the formulas below, \(\bar I\) denotes the complementary triple
of vector labels, and \(\alpha_I=s_Ic_{\bar I}-c_Is_{\bar I}\).

\subsection{Thermodynamics}
\label{sec:seven-thermodynamics}

On the \(\gamma=0\) subfamily, the compact duality rotations leave
the metric and its non-rotating AdS asymptotics unchanged. We follow
the thermodynamic normalization of the four-charge seed in
Ref.~\cite{Ma:2026fourcharge}. Take \(r_+\) to be the largest simple
zero of \(R_g\), with a regular exterior, in the conventions of
section~\ref{sec:eight-global}.
The entropy, temperature and angular velocity are
\begin{equation}
 S_+=\frac{\pi\sqrt{\mathcal F(r_+)}}{\Xi}\,,\qquad
 T_+=\frac{R_g'(r_+)}{4\pi\sqrt{\mathcal F(r_+)}}\,,\qquad
 \Omega_+=\frac{a\Xi L(r_+)}{\mathcal F(r_+)}\,.
\label{eq:seven-horizon-thermodynamics}
\end{equation}
With \(r_{I+}=r_++2ms_I^2\), the seed charges and their conjugate
chemical potentials have components
\begin{equation}
 Q_I^{(0)}=\frac{m c_Is_I}{2\sqrt\Xi}\,,\qquad
 \Phi_{I+}^{(0)}=\frac{2m}{\sqrt\Xi\,r_{I+}}
                     (c_Is_I-a\alpha_I\Omega_+)\,.
\label{eq:seven-seed-thermodynamics}
\end{equation}
The electric and magnetic charges and their chemical potentials are
\begin{equation}
\begin{aligned}
 \boldsymbol Q_{\rm el}&=\mathsf U\boldsymbol Q_{\rm el}^{(0)}\,,&
 \boldsymbol P&=\mathsf V\boldsymbol Q_{\rm el}^{(0)}\,,\\
 \boldsymbol\Phi_+&=\mathsf U\boldsymbol\Phi_+^{(0)}\,,&
 \boldsymbol\Psi_+&=\mathsf V\boldsymbol\Phi_+^{(0)}\,.
\end{aligned}
\label{eq:seven-charges-potentials}
\end{equation}
We retain the seed expressions for \(M\) and \(J\) as
\begin{equation}
 M=\frac{m}{\Xi}\left(1+\frac12\sum_{I=1}^4s_I^2\right)\,,
 \qquad
 J=\frac{m a}{\Xi}(c_{1234}-s_{1234})\,.
\label{eq:seven-mass-spin}
\end{equation}
In extended black-hole thermodynamics, the pressure and its
conjugate volume are
\begin{equation}
 P_{\rm th}=\frac{3g^2}{8\pi}\,,\qquad
 V_{\rm th}=\frac{\pi}{3\Xi}
                \left[\mathcal F'(r_+)+4a^2\Xi M\right]\,.
\label{eq:seven-pressure-volume}
\end{equation}

These expressions obey the following differential
identity on the solution parameter space, including variations of \(g\),
\begin{equation}
 \mathrm{d}M=T_+\,\mathrm{d}S_++\Omega_+\,\mathrm{d}J
       +\sum_{I=1}^4\left(\Phi_{I+}\,\mathrm{d}Q_I
                            +\Psi_{I+}\,\mathrm{d}P^I\right)
       +V_{\rm th}\,\mathrm{d}P_{\rm th}\,.
\label{eq:seven-first-law}
\end{equation}
Variations of the compact angles introduce no additional work term,
because the electric and magnetic contributions cancel.

This cancellation is an algebraic statement and does not imply that
angle variations preserve the boundary sources. In particular,
\eqref{eq:seven-boundary-gauge-curvature} shows that the boundary
electromagnetic curvature generally changes along the dyonic orbit.
A Hamiltonian interpretation requires an electromagnetic polarization,
a quantization and boundary condition for each scalar, and a choice
of finite boundary terms~\cite{Iyer:1994noether,Marolf:2006boundary,Lu:2013dyonic}.

For example, at fixed \(g\), an all-electric Dirichlet prescription
fixes the four boundary connections and all six leading scalar
coefficients in standard \(\Delta=2\) quantization. The scalar
coefficients may be evaluated in target-space normal coordinates
about the AdS vacuum. One may choose the Gibbons--Hawking term and
the corresponding divergent local counterterms, with no
electromagnetic Legendre term and no additional finite scalar,
mixed or boundary Chern--Simons terms. In this prescription the
magnetic fluxes are fixed, and admissible variations must preserve
all boundary sources. Generic boost and angle variations are
therefore restricted, and the seed expression for \(M\) cannot be
identified with the renormalized Hamiltonian energy without a
separate boundary-charge calculation. Equation~\eqref{eq:seven-first-law}
remains valid as a parameter-space identity, and its interpretation as
a first law on one fixed-source phase space requires the
corresponding integrability analysis.

\subsection{BPS limit}
\label{sec:seven-limits}

We take \(\gamma=0\) and the positive-charge, positive-rotation seed
branch, with \(\delta_I>0\).  On the FI-preserving duality orbit, 
the BPS relations and horizon potentials, in the seed normalization
of subsection~\ref{sec:seven-thermodynamics}, are~\footnote{The subscript ``\({\rm B}\)'' denotes BPS
values. In particular \(r_{\rm B}=\left.r_+\right|_{\rm B}\).}
\begin{equation}
\begin{aligned}
 T_{\rm B}&=0\,, & M_{\rm B}&=gJ_{\rm B}+\sum_{I=1}^4Q_{I\,,\rm B}\,,\\
 \Omega_{\rm B}&=g\,, & \Phi_{I\,,\rm B}&=1\,,\\
 \Psi_{I\,,\rm B}&=0\,, & \sum_{I=1}^4P_{\rm B}^I&=0\,.
\end{aligned}
\label{eq:seven-bps-thermodynamics}
\end{equation}
Writing
\begin{equation}
 d_{\rm B}=\sum_{I=1}^4\delta_I\,,\qquad
 \mathcal D_{\rm B}=\prod_{1\leq I<J\leq4}
                                   \sinh(\delta_I+\delta_J)\,.
\label{eq:seven-bps-auxiliary}
\end{equation}
the electric seed parameters are~\cite{Ma:2026fourcharge,Wu:2026fourcharge}
\begin{equation}
 m_{\rm B}=\frac{\sinh d_{\rm B}}
                   {2g\sqrt{\mathcal D_{\rm B}\cosh d_{\rm B}}}\,,
 \qquad
 a_{\rm B}=\frac1{g\cosh d_{\rm B}}\,,\qquad
 r_{\rm B}=\frac{2m_{\rm B}}{\sinh d_{\rm B}}
                         \sum_{I=1}^4c_Is_{\bar I}\,.
\label{eq:seven-bps-seed}
\end{equation}
Preserving the FI gauging restricts the duality angles to
\cite{Cacciatori:2016duality}
\begin{equation}
 \vartheta_2=\vartheta_1+\vartheta_3\,,\qquad
 (\psi_1\,,\psi_2\,,\psi_3\,,\psi_4)
      =(-2\vartheta_1\,,0\,,-2\vartheta_3\,,2\vartheta_1+2\vartheta_3)\,,
\label{eq:seven-bps-angles}
\end{equation}
with independent real angles \(\vartheta_1,\vartheta_3\).
The resulting BPS family generically carries six locally independent
classical electromagnetic charges~\cite{Hristov:2019dyonic}.

\subsection{Closed timelike curves}
\label{sec:seven-ctcs}

For \(\gamma=0\), excluding naked CTCs requires
\(g_{\phi\phi}>0\), or \(\mathcal H>0\)~\cite{Ma:2026fourcharge}.
On the regular exterior with \(m>0\) and \(r_I>0\),
\begin{equation}
\begin{aligned}
 \partial_r\mathcal H={}&
       \sum_{1\leq I<J<K\leq4}r_Ir_Jr_K
       +\frac{a^2}{2}(1+\cos^2\theta)\sum_{I=1}^4r_I\\
 &+m a^2\left(2+\sum_{I=1}^4s_I^2\right)\sin^2\theta>0\,.
\end{aligned}
\label{eq:seven-ctc-monotonicity}
\end{equation}
Hence
\begin{equation}
\begin{aligned}
 \mathcal H(r\,,\theta)
 &\geq\mathcal H(r_+\,,\theta)
\\
 &=\frac{\Delta_\theta}{\Xi}\mathcal F(r_+)\\
 &=\frac{\Delta_\theta}{\Xi}
       \left.W^2\right|_{r=r_+\,,\,\theta=0}>0\,.
\end{aligned}
\label{eq:seven-ctc-horizon-bound}
\end{equation}
Thus \(g_{\phi\phi}>0\) for \(r\geq r_+\), \(0<\theta<\pi\).

On the BPS branch, the Killing vector \(K_+\) of
section~\ref{sec:eight-ctcs} is non-spacelike. Its norm in
the non-extremal metric is
\begin{equation}
 K_+^2=-\frac{WR_g\Delta_\theta}{\Xi\mathcal H}
       +\frac{\mathcal H\sin^2\theta}{\Xi W}
          \left(\frac{a\Delta_\theta L}{\mathcal H}-g\right)^2\,.
\label{eq:seven-ctc-killing-norm}
\end{equation}
On the horizon, \(K_+^2\leq0\) and \(g_{\phi\phi}>0\) require
the squared term to vanish, as ensured by \(\Omega_{\rm B}=g\)
in \eqref{eq:seven-bps-thermodynamics}.

\subsection{Supersymmetric entropy function}
\label{sec:seven-entropy}

We restrict to the \(\gamma=0\) BPS sector of
section~\ref{sec:seven-limits}, at fixed duality angles satisfying
\eqref{eq:seven-bps-angles}, and suppress the labels \({\rm B}\) and
\(+\).\footnote{This constrained BPS functional of complex chemical
potentials differs from Sen's classical entropy-function
construction~\cite{Sen:2005entropy}. The quantum entropy function is
instead an \(\mathrm{AdS}_2\) path integral~\cite{Sen:2008quantum}.}
We use the \(k_{\mathrm{CS}}=1\) ABJM normalization for unquotiented
\(S^7\) and work at leading order in large \(N\) (classical
two-derivative supergravity). Here \(N\) is the M2-brane number and
\(g^{-2}=2\sqrt2N^{3/2}/3\), in the conventions of
section~\ref{sec:eight-global}. The charges entering the
supersymmetric entropy function~\cite{Choi:2018fdc} are\footnote{At fixed FI-preserving duality angles, the inverse
rotation in \eqref{eq:seven-entropy-charges} recovers the four electric
seed charges. The factor \(2/g\) converts them to the normalization
used in the entropy functional below.}
\begin{equation}
 \boldsymbol q=\frac2g
       \left(\mathsf U^T\boldsymbol Q_{\rm el}+\mathsf V^T\boldsymbol P\right)
       =\frac2g\boldsymbol Q_{\rm el}^{(0)}\,.
\label{eq:seven-entropy-charges}
\end{equation}
Equation~\eqref{eq:seven-entropy-charges} makes the functional below
the pullback of the electric-seed entropy functional under the inverse
duality rotation, at fixed FI-preserving angles. Its independent
charge arguments are the four \(q_I\). The two independent BPS
duality angles label the orbit on which this pullback is evaluated, and
they are not varied in the extremization.

For complex chemical potentials \(\Delta_I\),
\(I=1,\ldots,4\), and \(\omega\), 
the pulled-back electric-seed functional is
\begin{equation}
 \mathscr S(\boldsymbol\Delta\,,\omega;\boldsymbol q\,,J)
 =-\mathrm{i}\frac{4\sqrt2N^{3/2}}3
       \frac{\sqrt{\Delta_1\Delta_2\Delta_3\Delta_4}}{\omega}
       +\omega J+\sum_{I=1}^4\Delta_Iq_I\,,
\label{eq:seven-entropy-functional}
\end{equation}
subject to
\begin{equation}
 \sum_{I=1}^4\Delta_I-\omega=2\pi\mathrm{i}\,.
\label{eq:seven-entropy-constraint}
\end{equation}
At fixed angles, the Legendre term follows directly from
\eqref{eq:seven-entropy-charges}:
\[
 \sum_{I=1}^4\Delta_Iq_I
 =\frac2g\left[(\mathsf U\boldsymbol\Delta)^T\boldsymbol Q_{\rm el}
               +(\mathsf V\boldsymbol\Delta)^T\boldsymbol P\right].
\]
Thus the complex potentials are transported together with the charges.

With Lagrange multiplier \(\lambda\) for
\eqref{eq:seven-entropy-constraint}, extremizing
\(\mathscr S+\lambda(\sum_{I=1}^4\Delta_I-\omega-2\pi\mathrm{i})\)
gives the saddle equations
\begin{equation}
\begin{aligned}
 \lambda+q_I&=\mathrm{i}\frac{4\sqrt2N^{3/2}}{3\omega}
       \frac{\sqrt{\Delta_1\Delta_2\Delta_3\Delta_4}}{2\Delta_I}\,,
       \qquad I=1\,,2\,,3\,,4\,,\\
 \lambda-J&=\mathrm{i}\frac{4\sqrt2N^{3/2}}{3\omega^2}
                   \sqrt{\Delta_1\Delta_2\Delta_3\Delta_4}\,.
\end{aligned}
\label{eq:seven-entropy-extremum}
\end{equation}
At a saddle \((\star)\), with \(k_{\rm grav}=2N^3/9=1/(4g^4)\),
\begin{equation}
 S=\mathscr S_\star=-2\pi\mathrm{i}\lambda\,,
 \qquad
 \prod_{I=1}^4(\lambda+q_I)=-k_{\rm grav}(\lambda-J)^2\,.
\label{eq:seven-entropy-quartic}
\end{equation}
Writing the elementary symmetric polynomials as
\(\sigma_r=\sum_{1\leq I_1<\cdots<I_r\leq4}
q_{I_1}\cdots q_{I_r}\), \(r=1,2,3,4\), the positive entropy and
the associated charge relation are
\begin{equation}
\begin{aligned}
 S&=2\pi\sqrt{\frac{\sigma_3-2k_{\rm grav}J}{\sigma_1}}\,,\\
 0&=(\sigma_3-2k_{\rm grav}J)^2
    -\sigma_1(\sigma_2+k_{\rm grav})(\sigma_3-2k_{\rm grav}J)
    +\sigma_1^2(\sigma_4+k_{\rm grav}J^2)\,.
\end{aligned}
\label{eq:seven-entropy-positive}
\end{equation}
This form requires \(\sigma_1\ne0\) and
\((\sigma_3-2k_{\rm grav}J)/\sigma_1>0\). The physical root satisfies
\eqref{eq:seven-entropy-extremum} and
\eqref{eq:seven-entropy-constraint}, continuing the principal
large-black-hole branch along the regular BPS family.

We now verify the area matching explicitly, restoring BPS labels on
the metric parameters. Substituting \eqref{eq:seven-bps-seed} into
\eqref{eq:seven-parameters}, \eqref{eq:seven-seed-thermodynamics},
\eqref{eq:seven-mass-spin} and \eqref{eq:seven-entropy-charges} gives
\begin{equation}
\begin{aligned}
 \Xi_{\rm B}&=1-g^2a_{\rm B}^2=\tanh^2d_{\rm B}>0\,,\\
 q_I&=\frac2gQ_{I\,,\rm B}^{(0)}
     =\frac{m_{\rm B}s_Ic_I}{g\sqrt{\Xi_{\rm B}}}
     =\frac{s_Ic_I}{2g^2}
                  \sqrt{\frac{\cosh d_{\rm B}}{\mathcal D_{\rm B}}}\,,\\
 J&=\frac{m_{\rm B}a_{\rm B}}{\Xi_{\rm B}}
                         (c_{1234}-s_{1234})
   =\frac{c_{1234}-s_{1234}}{2g^2\sinh d_{\rm B}}
                  \sqrt{\frac{\cosh d_{\rm B}}{\mathcal D_{\rm B}}}\,.
\end{aligned}
\label{eq:seven-entropy-bps-substitution}
\end{equation}
For \(r=1,3\), write
\(\mathfrak e_r=\sum_{1\leq I_1<\cdots<I_r\leq4}
\prod_{p=1}^r s_{I_p}c_{I_p}\).
Expanding the hyperbolic addition formulae and using
\(c_I^2-s_I^2=1\) yields the identity
\begin{equation}
 \frac{\cosh d_{\rm B}\,\mathfrak e_3
       -2\mathcal D_{\rm B}(c_{1234}-s_{1234})/\sinh d_{\rm B}}
      {\mathfrak e_1}
 =s_{1234}+\frac{c_{1234}-s_{1234}}{\sinh d_{\rm B}}
                                     \sum_{I=1}^4c_Is_{\bar I}\,.
\label{eq:seven-entropy-boost-identity}
\end{equation}
The same BPS substitution in \eqref{eq:seven-metric-functions} gives
\begin{equation}
\begin{aligned}
 \left.\frac{g^4\mathcal F(r_+)}{\Xi^2}\right|_{\rm B}
 &=\frac{g^4}{\Xi_{\rm B}^2}
   \left[\left.r_{1234}\right|_{r=r_{\rm B}\,,\,m=m_{\rm B}}
           +a_{\rm B}^2\mathcal K(r_{\rm B})+a_{\rm B}^4\right]\\
 &=\frac1{\mathcal D_{\rm B}}\left[
 s_{1234}+\frac{c_{1234}-s_{1234}}{\sinh d_{\rm B}}
                                     \sum_{I=1}^4c_Is_{\bar I}\right]\,,
\end{aligned}
\label{eq:seven-entropy-area-identity}
\end{equation}
where all quantities on the right are restricted to the BPS seed.
Inserting \eqref{eq:seven-entropy-bps-substitution} into
\eqref{eq:seven-entropy-positive}, and then using
\eqref{eq:seven-entropy-boost-identity} and
\eqref{eq:seven-entropy-area-identity}, gives the continuous chain
\begin{equation}
\begin{aligned}
 \mathscr S_\star
 &=2\pi\sqrt{\frac{\sigma_3-2k_{\rm grav}J}{\sigma_1}}\\
 &=\frac{\pi}{g^2\sqrt{\mathcal D_{\rm B}}}
   \sqrt{\frac{\cosh d_{\rm B}\,\mathfrak e_3
       -2\mathcal D_{\rm B}(c_{1234}-s_{1234})/\sinh d_{\rm B}}
              {\mathfrak e_1}}\\
 &=\frac{\pi}{g^2\sqrt{\mathcal D_{\rm B}}}
   \sqrt{s_{1234}+\frac{c_{1234}-s_{1234}}{\sinh d_{\rm B}}
                                     \sum_{I=1}^4c_Is_{\bar I}}\\
 &=\left.\frac{\pi\sqrt{\mathcal F(r_+)}}{\Xi}\right|_{\rm B}
  =\left.S_+\right|_{\rm B}\,.
\end{aligned}
\label{eq:seven-entropy-area-matching}
\end{equation}
The second line includes \(k_{\rm grav}=1/(4g^4)\), while the third and
fourth lines use the two identities above. The last equality is the
horizon entropy \eqref{eq:seven-horizon-thermodynamics}. All square
roots take their positive values on the branch \(\delta_I>0\).
Since compact rotations leave the metric unchanged, for each fixed
FI-preserving duality rotation the pulled-back functional reproduces
the electric seed's area entropy.

The dyonic BPS entropy functional of Hristov, Katmadas and
Toldo~\cite{Hristov:2019dyonic} contains couplings between magnetic
charges and derivatives of the prepotential, as well as mixed electric
and magnetic contributions. Its purely electric sector
contains the seed functional used here, after translating conventions.
The present construction pulls back the electric-seed functional
for each chosen FI-preserving duality rotation. A direct comparison in physical dyonic variables
also requires matching the duality frame, charge and potential
normalizations, and boundary polarization. The area matching above
does not by itself derive an ABJM index for a general six-charge
fixed-source ensemble.

\section{Eight-charge spindle examples}
\label{app:eight-spindle-example}

This appendix gives extremal and non-extremal spindle branches of the
local solution in section~\ref{sec:seven-solution}. For coprime positive
integers \(k_\pm\), the polar opening angles \(2\pi/k_\pm\)\footnote{The resulting angular surface is an orbifold with underlying
topology \(S^2\) and local models \(\mathbb R^2/\mathbb Z_{k_\pm}\) at
the poles~\cite{Ferrero:2020twa}. Both four-dimensional axes are smooth
only when \(k_+=k_-=1\).} require
\begin{equation}
\begin{gathered}
 k_+^2\alpha_+=k_-^2\alpha_-\,,\qquad \alpha_\pm>0\,,\\
 \Delta\sigma_{\rm sp}=\frac{2\pi}{a k_+\sqrt{\alpha_+}}\,,\qquad
 \phi_{\rm sp}=k_+\phi\,,\qquad \Delta\phi_{\rm sp}=2\pi\,.
\end{gathered}
\label{eq:eight-extremal-spindle-quotient}
\end{equation}
The metric and matter fields are \eqref{eq:eight-metric},
\eqref{eq:seven-potentials} and \eqref{eq:eight-scalars}, with this
quotient. Relative to the north-regular normalization of
section~\ref{sec:eight-global}, the electromagnetic charges are
\begin{equation}
 \begin{pmatrix}\boldsymbol P_{\rm sp}\\
                 \boldsymbol Q_{{\rm el}\,,{\rm sp}}\end{pmatrix}
 =\frac1{k_+}
  \begin{pmatrix}\boldsymbol P\\\boldsymbol Q_{\rm el}\end{pmatrix}\,.
\label{eq:eight-spindle-fluxes}
\end{equation}

\subsection{An extremal example}
\label{app:eight-extremal-spindle-example}

At fixed \(g>0\), an extremal branch passes through
\begin{equation}
\begin{gathered}
 (e^{\delta_1}\,,e^{\delta_2}\,,e^{\delta_3}\,,e^{\delta_4}\,,e^\gamma)
       =(1\,,1\,,1\,,2\,,99/100)\,,\quad \vartheta_i=0\,,\quad
 (k_+\,,k_-)=(30000\,,29999)\,.
\end{gathered}
\label{eq:eight-special-extremal-branch}
\end{equation}
The bulk parameters \(m,a,r_{\rm e}\) satisfy
\eqref{eq:eight-extremal-spindle-quotient} and
\begin{equation}
\begin{gathered}
 R_g(r_{\rm e})=R_g'(r_{\rm e})=0\,,\qquad v_{\rm e}>0\,,\qquad
 \frac{1525}{1548}<\frac am<\frac{862}{875}\,,\qquad
 \frac9{10}<\frac{r_{\rm e}}m<1\,.
\end{gathered}
\label{eq:eight-special-extremal-parameters}
\end{equation}
The selected double root is the outer horizon, with the near-horizon
regularity conditions of section~\ref{sec:eight-extremal}. Its
temperature and entropy are
\begin{equation}
 T_{\rm sp}=0\,,\qquad
 S_{\rm sp}=\frac{S_{\rm e}}{30000}
 =\frac{\pi|L_{\rm e}|}{30000\sqrt{\alpha_+\beta_{\rm e}}}\,.
\label{eq:eight-extremal-spindle-flux-entropy}
\end{equation}

The smooth seven-charge extremal limit requires \(\gamma=0\) and
\(k_+=k_-=1\), which lies outside the fixed unequal-order branch. With
\(\Xi>0\), \(v_{\rm e}>0\) and a regular horizon and exterior,
the limiting results are
\begin{equation}
\begin{gathered}
 \Xi\Delta(r_{\rm e})+g^2\mathcal F(r_{\rm e})=0\,,\qquad
 \Xi(2r_{\rm e}-2m)+g^2\mathcal F'(r_{\rm e})=0\,,\qquad
 S_{\rm e}=\frac{\pi\sqrt{\mathcal F(r_{\rm e})}}{\Xi}\,.
\end{gathered}
\label{eq:eight-spindle-seven-extremality}
\end{equation}
The limiting fields and charges are those of
section~\ref{sec:seven-subfamily} and
\eqref{eq:seven-charges-potentials}. The additional BPS restrictions
are given in Appendix~\ref{sec:seven-limits}.

\subsection{A non-extremal example}
\label{app:eight-nonextremal-spindle-example}

A non-extremal branch passes through
\begin{equation}
\begin{gathered}
 m=1\,,\qquad a=\frac12\,,\qquad \vartheta_i=0\,,\\
 (e^{\delta_1}\,,e^{\delta_2}\,,e^{\delta_3}\,,e^{\delta_4}\,,e^\gamma)
 =\left(\frac32\,,\frac53\,,\frac74\,,\frac95\,,\frac34\right)\,,\\
 g^2=\frac{118451599865285100374016}
          {55971064860118724063218105}\,,\qquad
 (k_+\,,k_-)=(29999\,,30000)\,.
\end{gathered}
\label{eq:eight-nonextremal-spindle-parameters}
\end{equation}
The angular coefficients and outer horizon obey
\begin{equation}
 \frac{\alpha_-}{\alpha_+}
       =\left(\frac{29999}{30000}\right)^2\,,\qquad
 \alpha_+>0\,,\qquad
 \frac32<r_+<2\,,\qquad R_g'(r_+)>0\,.
\label{eq:eight-nonextremal-spindle-horizon}
\end{equation}
With the time normalization of \eqref{eq:eight-angular-identification}
and the azimuth \(\phi_{\rm sp}=29999\phi\), its horizon quantities are
\begin{equation}
 T_{\rm sp}=T_+>0\,,\qquad
 S_{\rm sp}=\frac{S_+}{29999}\,,\qquad
 \Omega_{\rm sp}=29999\,\Omega_+\,,
\label{eq:eight-nonextremal-spindle-data}
\end{equation}
where the right-hand sides are evaluated using
\eqref{eq:eight-horizon-data} at the same bulk parameters.

Near both specified points, the exterior satisfies the regularity and
causality conditions of section~\ref{sec:eight-ctcs} away from the
orbifold loci.

For the non-extremal branch, the fixed-order
parameterization, physical charge rank and exterior regularity are
established together in subsection~\ref{app:eight-fixed-spindle-family}.
Boundary sources may vary, and flux quantization is not imposed.

\subsection{A continuous family at fixed spindle orders}
\label{app:eight-fixed-spindle-family}

The non-extremal example in
\eqref{eq:eight-nonextremal-spindle-parameters} belongs to a continuous
eight-charge black-hole branch at fixed coupling and spindle orders.
For fixed \(g>0\) and coprime unequal integers \(k_+,k_->0\), introduce
\begin{equation}
\begin{aligned}
 A_{\rm sp}&:=\frac am>0\,,\qquad
 \widehat n:=\frac nm\,,\qquad
 \widehat v:=\frac{V_1}{m}\,,\qquad
 \eta_{\rm sp}:=\frac{k_+^2-k_-^2}{k_+^2+k_-^2}\,,\\
 \mathcal B_{\rm sp}&:=\eta_{\rm sp}
       (A_{\rm sp}^2+2\widehat n\widehat v-\widehat n^2)
                         +2A_{\rm sp}\widehat v\,.
\end{aligned}
\label{eq:spindle-continuous-parameters}
\end{equation}
The ratios \(\widehat n,\widehat v\) depend only on the five boosts.
The cone condition \eqref{eq:eight-extremal-spindle-quotient} fixes
the mass scale as
\begin{equation}
 m=\frac1g\sqrt{\frac{\eta_{\rm sp}}{\mathcal B_{\rm sp}}}\,,\qquad
 a=A_{\rm sp}m\,,
\label{eq:spindle-continuous-scale}
\end{equation}
and the angular coefficients become
\begin{equation}
 \alpha_\pm=\frac{2A_{\rm sp}\widehat v(1\mp\eta_{\rm sp})}
                          {\mathcal B_{\rm sp}}\,,\qquad
 \frac{\alpha_-}{\alpha_+}=\frac{k_+^2}{k_-^2}\,.
\label{eq:spindle-continuous-axes}
\end{equation}
This parameterization requires
\(\eta_{\rm sp}/\mathcal B_{\rm sp}>0\) and \(\alpha_\pm>0\).
It leaves nine continuous parameters: \(A_{\rm sp}\), the five boosts
\(\delta_I,\gamma\), and the three duality angles \(\vartheta_i\).
The fields are those of section~\ref{sec:seven-solution}, with the
quotient \eqref{eq:eight-extremal-spindle-quotient}.

For \((k_+,k_-)=(29999,30000)\), the charging data in
\eqref{eq:eight-nonextremal-spindle-parameters} and \(A_{\rm sp}=1/2\)
give a regular non-extremal solution for every prescribed \(g>0\)
through \eqref{eq:spindle-continuous-scale}. Its outer horizon is
the largest simple zero of \(R_g\), with
\begin{equation}
 \frac32<\frac{r_+}{m}<2\,,\qquad R_g'(r_+)>0\,.
\label{eq:spindle-regular-outer-horizon}
\end{equation}
Throughout \(r\geq r_+\) and \(u\in[n-a,n+a]\),
\begin{equation}
 W>0\,,\qquad N_i>0\quad(i=1\,,2\,,3)\,,\qquad
 1+g^2K>0\,,\qquad K+J_{\rm a}>0\,.
\label{eq:spindle-regular-exterior}
\end{equation}
The angular function satisfies \(1-g^2J_{\rm a}>0\), and
\(R_g>0\) for \(r>r_+\). These inequalities establish a real scalar
patch and the sufficient no-CTC conditions of
section~\ref{sec:eight-ctcs}, away from the prescribed orbifold loci.

At this same solution, denoted by \(*\), the physical spindle fluxes
in \eqref{eq:eight-spindle-fluxes} satisfy
\begin{equation}
 \operatorname{rank}\left[\left.
 \frac{\partial(\boldsymbol Q_{{\rm el}\,,{\rm sp}}\,,\boldsymbol P_{\rm sp})}
 {\partial(\delta_1\,,\delta_2\,,\delta_3\,,\delta_4\,,\gamma\,,
                   \vartheta_1\,,\vartheta_2\,,\vartheta_3)}
 \right|_{*\,,\,g\,,k_+\,,k_-\,,A_{\rm sp}}\right]=8\,.
\label{eq:spindle-physical-charge-rank}
\end{equation}
This Jacobian includes the boost dependence of the mass scale and
flux normalization. Thus the eight electromagnetic charges are locally
independent even at fixed \(A_{\rm sp}\).
The strict exterior inequalities and the nonzero Jacobian determinant
persist in a sufficiently small neighbourhood, giving an open
non-extremal eight-charge black-hole branch at these fixed spindle
orders. Other unequal orders are covered by the same parameterization,
subject to separate horizon and exterior regularity conditions.
The charge count is classical and allows the boundary sources to vary.

\section{Black hole entropy and soft hair}
\label{app:Phase space formalism}

The Kerr/CFT correspondence was proposed as a duality between
extremal Kerr and a chiral two-dimensional conformal field theory,
based on the asymptotic symmetries of its near-horizon
geometry~\cite{Guica:2008mu}. For generic non-extremal Kerr, a hidden
conformal symmetry was subsequently found in the low-frequency scalar
wave equation in the near region~\cite{Castro:2010fd}. This structure
motivated a horizon symmetry construction whose surface charges and
central terms are evaluated using the covariant phase space formalism
with a boundary counterterm~\cite{Haco:2018ske}. Assuming a quantum
realization of the resulting Virasoro symmetries and the applicability
of the Cardy formula, the construction reproduces the Kerr area
entropy. Extensions include Kerr--Newman~\cite{Haco:2019ggi},
Kerr--AdS~\cite{Perry:2020ndy}, and Kerr--Taub--NUT and more general
four-dimensional geometries~\cite{Perry:2022udk}.
In the soft-hair interpretation, these horizon symmetries describe
gravitational degrees of freedom associated with large diffeomorphisms
carrying nontrivial horizon surface
charges~\cite{Haco:2018ske,Haco:2019ggi,Perry:2020ndy}.

We apply this construction to the full eight-charge metric
\eqref{eq:eight-metric}, using the north-regular quotient and time
normalization of \eqref{eq:eight-angular-identification}. The boundary
sources and axial defect are held fixed throughout the horizon
analysis.
This does not by itself fix the polar-tube boundary conditions or
establish global charge integrability; these requirements are discussed
in subsection~\ref{app:covariant-charges}. 
With the normalization specified below, the thermal parameters and
evaluated local central coefficients reproduce the area expressions
algebraically, and a Cardy interpretation requires the stated phase-space
and quantum assumptions.
For the eight-charge family these are regulated local
horizon data of the complete local fields in
section~\ref{sec:seven-solution}. Their restriction to the complete
smooth-axis seven-charge solution is collected in
subsection~\ref{app:seven-complete-horizon-results}.

\subsection{Horizon geometry}
\label{app:eight-conformal-metric}

We consider adjacent simple horizons \(r_-<r_+\), with
\(R_g(r_\pm)=0\), on a real patch with positive horizon areas.
Using \(u=n+a\cos\theta\), \(0<\theta<\pi\), and
\(\Upsilon(\theta)=1-g^2J_{\rm a}(n+a\cos\theta)\),
the normalized metric is\footnote{\(\Omega_{\mathrm{drag}}\) is the local
frame-dragging angular velocity
\[
 \Omega_{\mathrm{drag}}
 =-\frac{g_{t\phi}}{g_{\phi\phi}}
 =\frac{a\Upsilon L}{\mathcal H}\,.
\]}
\begin{equation}
\begin{aligned}
 \mathrm ds_4^2&=
 -\frac{WR_g\Upsilon}{\alpha_+\mathcal H}\,\mathrm dt^2
 +\frac{\mathcal H\sin^2\theta}{\alpha_+W}
            (\mathrm d\phi-\Omega_{\mathrm{drag}}(r\,,\theta)\mathrm dt)^2
 +W\left(\frac{\mathrm dr^2}{R_g}
                         +\frac{\mathrm d\theta^2}{\Upsilon}\right)\,,\\
 \Omega_{\mathrm{drag}}(r\,,\theta)&=\frac{a\Upsilon L}{\mathcal H}\,.
\end{aligned}
\label{eq:appC-rotating-metric}
\end{equation}
Here \(W\) and \(\mathcal H\) are evaluated at \(u=n+a\cos\theta\), and \(\phi\)
has period \(2\pi\). We require \(\alpha_+>0\), \(W>0\) and
\(\Upsilon>0\) on the patch under consideration. Together with the
outer-horizon coefficients \eqref{eq:eight-horizon-coefficients}, we use
\begin{equation}
\begin{aligned}
 \beta_-&=1+g^2K(r_-)>0\,, & L_-&=L(r_-)\ne0\,,\\
 \mathcal A_\pm&=\frac{\sqrt{\alpha_+}|L_\pm|}{\sqrt{\beta_\pm}}\,, & D_+&=R_g'(r_+)>0\,.
\end{aligned}
\label{eq:appC-horizon-data}
\end{equation}
The temperatures, angular velocities and entropies then read\footnote{The inner-horizon temperature retains the sign of
\(R_g'(r_-)\), which is negative for the adjacent simple horizons
considered here. Replacing it by \(|T_-|\) requires corresponding sign
changes in inner-horizon thermodynamic relations.}
\begin{equation}
\begin{aligned}
 \left.\mathcal H\right|_{r_\pm}
      &=\frac{\Upsilon L_\pm^2}{\beta_\pm}
      =\frac{\Upsilon\mathcal A_\pm^2}{\alpha_+}\,,\\
 T_\pm&=\frac{R_g'(r_\pm)}{4\pi\mathcal A_\pm}\,,\\
 \Omega_\pm&=\frac{a\beta_\pm}{L_\pm}\,,\\
 S_\pm&=\frac{\pi\mathcal A_\pm}{\alpha_+}\,.
\end{aligned}
\label{eq:appC-thermodynamics}
\end{equation}
We work on the patch
\(\mathcal A_+>\mathcal A_->0\), \(\Omega_->\Omega_+\).
These quantities depend on the full five-boost metric and require
no identification of its parameters with conserved mass or angular
momentum.

\subsection{Conformal coordinates}
\label{app:conformal-coordinates}

The real coefficients \(a_R,b_R,a_L,b_L\) are distinct from the
rotation, charging boosts and duality angles. Introduce right and
left coordinates
\begin{equation}
 t_R={a_R}\phi+{b_R} t\,,\qquad
 t_L={a_L}\phi+{b_L} t\,,
\label{eq:appC-tRL}
\end{equation}
We impose the
outer-horizon normalization
\(a_R\Omega_++b_R=D_+/(2\mathcal A_+)\) 
and choose the remaining relative
normalization so that a common local central coefficient is compatible
with both horizon areas. The coefficients are
\begin{equation}
\begin{aligned}
 {a_R}&=\frac{{D_+}(\mathcal A_+-\mathcal A_-)}
          {2\mathcal A_+\mathcal A_-(\Omega_--\Omega_+)}\,,&
 {a_L}&=\frac{{D_+}(\mathcal A_++\mathcal A_-)}
          {2\mathcal A_+\mathcal A_-(\Omega_--\Omega_+)}\,,\\
 {b_R}&=\frac {D_+}{2\mathcal A_+}-{a_R}\Omega_+\,,&
 {b_L}&=-\frac {D_+}{2\mathcal A_+}-{a_L}\Omega_+\,.
\end{aligned}
\label{eq:appC-coordinate-parameters}
\end{equation}
A convenient radial choice is
\begin{equation}
 \mathcal R(r)^2=\frac{(r_+-r_-)(r-r_+)}{{D_+}(r-r_-)}\,,\qquad
 \mathcal Q_{\rm conf}(r)^2=1-\mathcal R(r)^2\,,
\label{eq:appC-radial-coordinates}
\end{equation}
with positive branches in an exterior neighbourhood of \(r_+\).
The conformal coordinates are
\begin{equation}
 w^+=\mathcal R(r)e^{t_R}\,,\qquad
 w^-=\mathcal R(r)e^{t_L}\,,\qquad
 y=\mathcal Q_{\rm conf}(r)e^{(t_L+t_R)/2}\,.
\label{eq:appC-conformal-map}
\end{equation}
The outer bifurcation surface lies at
\(w^+=w^-=0\).

The radial angular-velocity gradient
\(\omega_1(\theta)=
\left.\partial_r\Omega_{\mathrm{drag}}(r,\theta)\right|_{r_+}\)
is
\begin{equation}
\begin{aligned}
 \omega_1(\theta)=\frac{a\beta_+}{L_+}\Bigg[
 \frac{L_1}{L_+}
 -\frac{\beta_+}{\Upsilon(\theta)L_+^2}
 \Big\{&(2r_++4\mathfrak m-2m)R_{\rm s}(r_+)\\
 &+2(r_+-m)\bigl[K(r_+)+J_{\rm a}(n+a\cos\theta)\bigr]
 +2L_+L_1\Big\}\Bigg]\,.
\end{aligned}
\label{eq:appD-eight-drag-gradient}
\end{equation}
Expanding the metric coefficients in \(w^\pm\) about the bifurcation
surface, with \(y\) and \(\theta\) held fixed, gives
\begin{equation}
\begin{aligned}
 \mathrm ds_4^2\simeq{}&
 \frac{4W_+}{y^2}\,\mathrm dw^+\mathrm dw^-
 +\frac{\mathfrak k_H^2\Upsilon\sin^2\theta}{W_+y^2}\,
       \mathrm dy^2
 +\frac{W_+}{\Upsilon}\,\mathrm d\theta^2\\
 &+\frac{4w^+}{y^3}
 \left[
 \frac{\mathfrak k_H^2\Upsilon\sin^2\theta}{4W_+}
 +\frac{\Upsilon\mathcal A_+^3\sin^2\theta}
        {\alpha_+^2W_+(a_R+a_L)}\,\omega_1
 -\frac{2a_RW_+}{a_R+a_L}
 \right]\mathrm dw^-\mathrm dy\\
 &+\frac{4w^-}{y^3}
 \left[
 \frac{\mathfrak k_H^2\Upsilon\sin^2\theta}{4W_+}
 -\frac{\Upsilon\mathcal A_+^3\sin^2\theta}
        {\alpha_+^2W_+(a_R+a_L)}\,\omega_1
 -\frac{2a_LW_+}{a_R+a_L}
 \right]\mathrm dw^+\mathrm dy\,,
\end{aligned}
\label{eq:appD-eight-bifurcation-metric}
\end{equation}
where 
\begin{align}
      W_+(\theta)=W(r_+\,,n+a\cos\theta)\,,\quad 
\mathfrak k_H=-\frac{2\mathcal A_+}{\alpha_+(a_R+a_L)}\,.
\end{align}
Here \(W_+,\Upsilon,\omega_1\) are evaluated at the same \(\theta\)\footnote{The symbol \(\simeq\) retains the constant and linear terms in
\(w^\pm\) in the metric coefficients. }.
The omitted coefficients
start at total degree two. In particular,
\eqref{eq:appC-radial-coordinates} gives
\(r-r_+=D_+w^+w^-/y^2+O((w^+w^-)^2)\).
Both linear mixed terms are retained for the horizon charge
calculation. The seven-charge restriction is obtained by setting
\(\gamma=0\), \(\alpha_+=\Xi\) and
\(\Upsilon=\Delta_\theta\).

Under \(\phi\sim\phi+2\pi\),
\begin{equation}
 w^+\sim e^{2\pi{a_R}}w^+\,,\qquad
 w^-\sim e^{2\pi{a_L}}w^-\,,\qquad
 y\sim e^{\pi({a_R}+{a_L})}y\,.
\label{eq:appC-identifications}
\end{equation}
The associated thermal parameters \(T_R=a_R/(2\pi)\) and
\(T_L=a_L/(2\pi)\) are therefore
\begin{equation}
 T_R=\frac{{D_+}(\mathcal A_+-\mathcal A_-)}
       {4\pi\mathcal A_+\mathcal A_-(\Omega_--\Omega_+)}\,,\qquad
 T_L=\frac{{D_+}(\mathcal A_++\mathcal A_-)}
       {4\pi\mathcal A_+\mathcal A_-(\Omega_--\Omega_+)}\,.
\label{eq:appC-temperatures}
\end{equation}

\subsection{Conformal vector fields}
\label{app:conformal-vectors}

With \(\partial_\pm=\partial/\partial w^\pm\), the modes
compatible with \eqref{eq:appC-identifications} are
\begin{equation}
\begin{aligned}
 \zeta_\ell&=(w^+)^{\mathrm{i} \ell/a_R}
       \left[a_Rw^+\partial_+
                    +\tfrac12(a_R+\mathrm{i} \ell)y\partial_y\right]\,,\\
 \bar\zeta_\ell&=(w^-)^{\mathrm{i} \ell/a_L}
       \left[a_Lw^-\partial_-
                    +\tfrac12(a_L+\mathrm{i} \ell)y\partial_y\right]\,,
 \qquad \ell\in\mathbb Z\,.
\end{aligned}
\label{eq:appC-conformal-vectors}
\end{equation}
For integer mode labels \(\ell,p\), the brackets are
\[
\begin{aligned}
 [\zeta_\ell,\zeta_p]&=\mathrm{i}(p-\ell)\zeta_{\ell+p},&
 [\bar\zeta_\ell,\bar\zeta_p]&=\mathrm{i}(p-\ell)\bar\zeta_{\ell+p},\\
 [\zeta_\ell,\bar\zeta_p]&=0.
\end{aligned}
\]
These local near-horizon vector fields form two commuting Witt
algebras. The right- and left-moving modes preserve the future and
past horizons, respectively. Their extensions away from the horizon
are chosen to vanish near the conformal boundary; the displayed
brackets specify the local near-horizon algebra.

\subsection{Covariant charges}
\label{app:covariant-charges}

We use the covariant prescription
\cite{Lee:1990nz,Iyer:1994noether,Wald:1999wa} with the
normal-bundle counterterm of
Refs.~\cite{Perry:2020ndy,Perry:2022udk}. Horizon integrals are
evaluated with polar caps excised and the cap size taken to zero.
The result gives the local horizon contribution.
A global completion at polar tubes and the defect core is not
specified here. For gauge-covariant
generators in regular horizon patches, the Maxwell and scalar
contributions to the cubic central term vanish~\cite{Haco:2019ggi}.

For a regulated horizon cut \(\mathcal S\) and a combined generator
\(\varepsilon=(\zeta,\lambda^I)\), the linearized covariant charges
are specified by
\begin{equation}
\begin{aligned}
 \mathcal Q_\varepsilon
 &=-\frac1{16\pi}\left[\star\mathrm d\zeta^\flat
       +(\iota_\zeta B_{(1)}^I+\lambda^I)G_I\right]\,,\\
 \delta H_\varepsilon[\mathcal S]
 &=\int_{\mathcal S}\left[
   \delta\mathcal Q_\varepsilon-\mathcal Q_{\delta\varepsilon}
   -\iota_\zeta\Theta_{\rm ps}+\frac1{16\pi}\star F_{\rm ct}\right]\,,
 \qquad \lambda^I=-\iota_\zeta B_{(1)}^I \,.
\end{aligned}
\label{eq:appD-covariant-horizon-charges}
\end{equation}
Here \(\Theta_{\rm ps}\) is the presymplectic potential of the
all-electric STU action with overall normalization \(1/(16\pi)\).
The inverse-metric perturbation is \(\widehat h^{ab}=\delta g^{ab}\).
Here \(a,b,c,d\) are four-dimensional spacetime indices.
The horizon binormal \(N_{ab}\) is oriented by
\(N_+{}^+=1\) and \(N_-{}^-=-1\).
The normal-bundle prescription is
\begin{equation}
\begin{aligned}
 F_{\rm ct}&=\frac12F_{{\rm ct}\,ab}\,
                         \mathrm dx^a\wedge\mathrm dx^b\,,\\
 F_{{\rm ct}\,ab}
 &=-4N_d{}^c\nabla_c(\zeta_{[a}\widehat h^d{}_{b]})\\
 &=-2N_d{}^c\left[
   \nabla_c(\zeta_a\widehat h^d{}_b)
  -\nabla_c(\zeta_b\widehat h^d{}_a)\right]\,.
\end{aligned}
\label{eq:appD-horizon-counterterm}
\end{equation}
The subtraction \(\mathcal Q_{\delta\varepsilon}\) accounts
for the field dependence of the compensating gauge parameter.
For horizon modes we use a field-dependent gauge-covariant lift.
For \(\zeta=\xi_+\), this choice agrees on the horizon with the constant
polar-patch compensators in \eqref{eq:eight-horizon-gauge-lift}. Away
from the horizon it generally has a different extension.
Writing \(F_{\rm IW}=\frac12F_{{\rm IW}\,ab}\,
\mathrm dx^a\wedge\mathrm dx^b\) and
\(\widehat h=g_{ab}\widehat h^{ab}\), the explicit Iyer--Wald
two-form in the same convention is
\begin{equation}
\begin{aligned}
 F_{{\rm IW}\,ab}={}&
 \frac12\nabla_a\zeta_b\,\widehat h
 +\nabla_a\widehat h^c{}_b\,\zeta_c
 +\nabla_c\zeta_a\,\widehat h^c{}_b\\
 &+\nabla_c\widehat h^c{}_a\,\zeta_b
 -\nabla_a\widehat h\,\zeta_b
 -(a\leftrightarrow b)\,.
\end{aligned}
\label{eq:appD-Iyer-Wald-two-form}
\end{equation}
The gravitational horizon variation therefore splits into
\begin{equation}
\begin{aligned}
 \delta H_\zeta^{\rm grav}
 &=\delta H_\zeta^{\rm IW}+\delta H_\zeta^{\rm ct}\,,\\
 \delta H_\zeta^{\rm IW}
 &=\frac1{16\pi}\int_{\mathcal S}\star F_{\rm IW}\,,
 \qquad
 \delta H_\zeta^{\rm ct}
 =\frac1{16\pi}\int_{\mathcal S}\star F_{\rm ct}\,.
\end{aligned}
\label{eq:appD-IW-counterterm-split}
\end{equation}
For a mode variation generated by a second vector field \(\tilde\zeta\),
\(\widehat h^{ab}=\mathcal L_{\tilde\zeta}g^{ab}\).
The covariant metric variation is
\(\delta g_{ab}=-g_{ac}g_{bd}\widehat h^{cd}\).
The right- and left-moving charge variations are obtained by taking
\(\zeta=\zeta_\ell\) and \(\zeta=\bar\zeta_\ell\), respectively.
For these variations, the scalar and electromagnetic surface forms
vanish on the horizon interior. At fixed fluxes the north--south
transition functions are preserved by the gauge-covariant lift,
and there is no electromagnetic patch-seam contribution.

On the black-hole background, the local central terms
for the right- and left-moving modes \(\zeta_\ell\) and
\(\bar\zeta_\ell\), respectively, evaluated with the same
surface-orientation convention, are
\begin{equation}
\begin{aligned}
 K_{\ell\,,p}^{(R)}&=\frac{\mathrm{i}\mathcal A_+}
                  {2\alpha_+(a_R+a_L)}
               (\ell^3+a_R^2\ell)\delta_{\ell+p\,,0}\,,\\
 K_{\ell\,,p}^{(L)}&
 =\frac{\mathrm{i}\mathcal A_+}
                  {2\alpha_+(a_R+a_L)}
               (\ell^3+a_L^2\ell)\delta_{\ell+p\,,0}\,.
\end{aligned}
\label{eq:appC-central-term}
\end{equation}
After shifting the zero modes to remove the linear terms, the
local right- and left-moving central coefficients are
\begin{equation}
 c_L=c_R=
 \frac{6\mathcal A_+\mathcal A_-(\Omega_--\Omega_+)}
        {\alpha_+D_+}\,.
\label{eq:appC-central-charges}
\end{equation}
The complex-mode convention \eqref{eq:appC-conformal-vectors}
uses the Dirac-to-commutator factor \(-\mathrm{i}\)
\cite{Perry:2020ndy}. These are central terms at a chosen background,
while a Virasoro charge-algebra interpretation requires an integrable reduced phase
space with constant central charges.
For polar caps of angular size \(\delta\), the
omitted horizon disks give \(O(\delta^2)\). The nonzero modes do not
extend smoothly to the poles, including on the smooth seven-charge
background, so polar-tube boundary conditions and global charge
integrability remain additional requirements. The asymptotic
eight-charge fluxes are those of \eqref{eq:eight-fluxes}. 
For the general eight-charge branch, an integrable Hamiltonian mass
and angular momentum have not been established, as explained in
subsection~\ref{sec:eight-horizon}.

\subsection{The area law}
\label{app:area-law}

With the coordinate normalization in
subsection~\ref{app:conformal-coordinates}, the Cardy expression gives
the algebraic area matching
\begin{equation}
 S_\pm=\frac{\pi^2}{3}(c_LT_L\pm c_RT_R)
      =\frac{\pi\mathcal A_\pm}{\alpha_+}\,.
\label{eq:appC-Cardy}
\end{equation}
These expressions are independent of the compact duality angles.
At \(\gamma=0\), \(\alpha_+=\Xi\), \(\Upsilon=\Delta_\theta\) and
\(\mathcal A_\pm=\sqrt{\mathcal F(r_\pm)}\), recovering the
seven-charge results. The temperatures and central charges are
invariant under constant azimuthal-frame changes, which shift both
\(\Omega_\pm\) equally.
This agreement is a consistency check of the chosen thermal
normalization. A soft-hair interpretation additionally requires a
globally integrable horizon phase space, a quantum realization of its
Virasoro algebra and a regime in which the Cardy formula applies. Establishing its quantum state space and a direct microscopic
counting remains an open problem.

\subsection{Results for the complete seven-charge solution}
\label{app:seven-complete-horizon-results}

The restriction \(\gamma=0\) gives the complete seven-charge metric
and matter fields of Appendix~\ref{app:seven-subfamily}, with
\(n=V_1=0\), \(\alpha_+=\alpha_-=\Xi>0\) and two smooth axes~\footnote{A superscript \((7)\) below indicates this restriction.}.

On the two-horizon patch used above,
\begin{equation}
 \mathcal A_\pm=\sqrt{\mathcal F(r_\pm)}\,,\qquad
 D_+=R_g'(r_+)\,,\qquad
 \Omega_\pm=\frac{a\Xi L(r_\pm)}{\mathcal F(r_\pm)}\,.
\label{eq:appD-seven-horizon-restriction}
\end{equation}
The asymptotic electromagnetic fluxes and the seed-normalized mass
and angular momentum reduce to
\begin{equation}
\begin{aligned}
 Q_I^{(0)}&=\frac{mc_Is_I}{2\sqrt\Xi}\,,&
 \boldsymbol Q_{\rm el}&=\mathsf U\boldsymbol Q_{\rm el}^{(0)}\,,&
 \boldsymbol P&=\mathsf V\boldsymbol Q_{\rm el}^{(0)}\,,\\
 M&=\frac m\Xi\left(1+\frac12\sum_{I=1}^4s_I^2\right)\,,&
 J&=\frac{ma}{\Xi}(c_{1234}-s_{1234})\,.
\end{aligned}
\label{eq:appD-seven-bulk-charge-data}
\end{equation}
The boundary prescription and Hamiltonian
interpretation of \(M,J\) are those discussed in
subsection~\ref{sec:seven-thermodynamics}.

The right- and left-moving temperatures and local central
coefficients are
\begin{equation}
\begin{aligned}
 T_R^{(7)}&=\frac{D_+(\mathcal A_+-\mathcal A_-)}
 {4\pi\mathcal A_+\mathcal A_-(\Omega_--\Omega_+)}\,,\\
 T_L^{(7)}&=\frac{D_+(\mathcal A_++\mathcal A_-)}
 {4\pi\mathcal A_+\mathcal A_-(\Omega_--\Omega_+)}\,,\\
 c_L^{(7)}&=c_R^{(7)}
 =\frac{6\mathcal A_+\mathcal A_-(\Omega_--\Omega_+)}
 {\Xi D_+}\,.
\end{aligned}
\label{eq:appD-seven-cft-results}
\end{equation}
The corresponding linearized horizon charges and their local
central terms are the \(\gamma=0\) restrictions of
\eqref{eq:appD-covariant-horizon-charges} and
\eqref{eq:appC-central-term}.
With the same normalization, the Cardy expression gives the
outer-horizon area entropy and inner-horizon area combination
algebraically as
\begin{equation}
 S_\pm^{(7)}
 =\frac{\pi^2}{3}\left(c_L^{(7)}T_L^{(7)}
                     \pm c_R^{(7)}T_R^{(7)}\right)
 =\frac{\pi\mathcal A_\pm}{\Xi}
 =\frac{\pi\sqrt{\mathcal F(r_\pm)}}{\Xi}\,.
\label{eq:appD-seven-entropy-results}
\end{equation}
These geometric results are independent of the three compact
duality angles. Smoothness of the seven-charge background removes
the physical conical core. The polar regularity and charge-integrability
requirements on the nonzero horizon modes remain those stated above.

\section{Higher-dimensional comparison tables}
\label{app:higher-dimensional-comparison}

Tables~\ref{tab:ads5-solutions} and~\ref{tab:ads7-solutions} summarize
representative charged AdS black holes in five and seven dimensions.
The charge labels refer to electric charges in the indicated gauged
supergravity truncations.

\subsection{Five-dimensional solutions}

For asymptotically global AdS\(_5\) black holes with spherical horizons
and vanishing boundary magnetic fields, the asymptotic \(S^3\) has no
nontrivial two-cycles supporting magnetic monopole flux. The Hodge
dual of a two-form is a three-form, so electric--magnetic rotations
do not preserve the vector-field content. Magnetically charged black
strings~\cite{Hosseini:2016cyf} and black holes supported by boundary
magnetic fields~\cite{BlazquezSalcedo:2017magnetized} belong to
different sectors.

\begin{table}[!htbp]
\centering
\small
\setlength{\tabcolsep}{7pt}
\setlength{\arrayrulewidth}{0.4pt}
\renewcommand{\arraystretch}{1.45}
\caption{Exact AdS\(_5\) black holes in gauged STU supergravity and its
truncations.}
\label{tab:ads5-solutions}
\begin{tabularx}{\textwidth}{|>{\raggedright\arraybackslash}X|>{\centering\arraybackslash}p{0.23\textwidth}|}
\hline
Black hole parameters & References \\
\hline
Static: \(Q_I;\ a_1=a_2=0\)
& \cite{Behrndt:1999static5d} \\
\hline
Equal rotations: \(Q_I;\ a_1=a_2\)
& \cite{Cvetic:2004threecharge} \\
\hline
One charging parameter: \(Q_1=Q_2,\ Q_3\ \text{induced};\ a_1,a_2\)
& \cite{Chong:2005da} \\
\hline
Minimal truncation: \(Q_1=Q_2=Q_3;\ a_1,a_2\)
& \cite{Chong:2005hr} \\
\hline
Single electric charge: \(Q_1,\ Q_2=Q_3=0;\ a_1,a_2\)
& \cite{Chong:2006zx} \\
\hline
General non-extremal: \(Q_I;\ a_1,a_2\)
& \cite{Wu:2012general5d} \\
\hline
BPS: \(Q_I;\ J_1,J_2\)
& \cite{Kunduri:2006ek} \\
\hline
\end{tabularx}
\par\vspace{4pt}
\begin{minipage}{\textwidth}
\footnotesize
\textbf{Notation}: \(Q_I\) (\(I=1,2,3\)) are electric charges;
\(a_i\) and \(J_i\) (\(i=1,2\)) are rotation parameters and
angular momenta.
An independent mass is implicit outside the BPS row.
In the row with one charging parameter, \(Q_3\) is induced by the
same charging parameter and the rotations.
The five BPS conserved quantities obey one relation.
\end{minipage}
\end{table}

\clearpage

\subsection{Seven-dimensional solutions}

The \(U(1)^2\) Cartan truncation of seven-dimensional \(SO(5)\)
gauged supergravity contains two gauge fields, two real scalars and
a three-form potential subject to odd-dimensional
self-duality~\cite{Cvetic:1999xp,Liu:1999ads7,Chong:2004dy}.
The table includes the general non-extremal family and its BPS
limit~\cite{Bobev:2023ads7}.

\begin{table}[!htbp]
\centering
\small
\setlength{\tabcolsep}{7pt}
\setlength{\arrayrulewidth}{0.4pt}
\renewcommand{\arraystretch}{1.45}
\caption{Exact AdS\(_7\) black holes in \(U(1)^2\) gauged
supergravity with spherical horizons.}
\label{tab:ads7-solutions}
\begin{tabularx}{\textwidth}{|>{\raggedright\arraybackslash}X|>{\centering\arraybackslash}p{0.23\textwidth}|}
\hline
Black hole parameters & References \\
\hline
Static: \(Q_I;\ a_1=a_2=a_3=0\)
& \cite{Cvetic:1999xp,Liu:1999ads7} \\
\hline
Equal rotations: \(Q_I;\ a_1=a_2=a_3\)
& \cite{Chong:2004dy} \\
\hline
Equal charges: \(Q_1=Q_2;\ a_1,a_2,a_3\)
& \cite{Chow:2007ts} \\
\hline
Single rotation: \(Q_I;\ a_1,\ a_2=a_3=0\)
& \cite{Wu:2011ads7,Chow:2011singleRotation} \\
\hline
General non-extremal: \(Q_I;\ a_1,a_2,a_3\)
& \cite{Bobev:2023ads7} \\
\hline
BPS: \(Q_I;\ J_1,J_2,J_3\)
& \cite{Chow:2007ts,Bobev:2023ads7} \\
\hline
\end{tabularx}
\par\vspace{4pt}
\begin{minipage}{\textwidth}
\footnotesize
\textbf{Notation}: \(Q_I\) (\(I=1,2\)) are electric charges;
\(a_i\) and \(J_i\) (\(i=1,2,3\)) are rotation parameters and
angular momenta. An independent mass is implicit outside the BPS
row. For regular BPS black holes, supersymmetry fixes the mass and
the five displayed conserved quantities obey one nonlinear relation.
\end{minipage}
\end{table}

\FloatBarrier

\bibliographystyle{utphys}
\bibliography{ref}

@book{Stephani2003,
  author        = {Stephani, Hans and Kramer, Dietrich and MacCallum, Malcolm and Hoenselaers, Cornelius and Herlt, Eduard},
  title         = {Exact Solutions of {Einstein}'s Field Equations},
  year          = {2003},
  edition       = {2},
  series        = {Cambridge Monographs on Mathematical Physics},
  publisher     = {Cambridge University Press},
  doi           = {10.1017/CBO9780511535185}
}

@article{Ren:2019lgw,
  author        = {Ren, Jie},
  title         = {Analytic solutions of neutral hyperbolic black holes with scalar hair},
  journal       = {Phys. Rev. D},
  volume        = {106},
  number        = {8},
  pages         = {086023},
  year          = {2022},
  doi           = {10.1103/PhysRevD.106.086023},
  eprint        = {1910.06344},
  archivePrefix = {arXiv},
  primaryClass  = {hep-th}
}

@article{Lu:2014sza,
  author        = {L{\"u}, H. and V{\'a}zquez-Poritz, Justin F.},
  title         = {C-metrics in Gauged {STU} Supergravity and Beyond},
  journal       = {JHEP},
  volume        = {12},
  pages         = {057},
  year          = {2014},
  doi           = {10.1007/JHEP12(2014)057},
  eprint        = {1408.6531},
  archivePrefix = {arXiv},
  primaryClass  = {hep-th}
}

@article{Ferrero:2020twa,
  author        = {Ferrero, Pietro and Gauntlett, Jerome P. and Ipi{\~n}a, Juan Manuel P{\'e}rez and Martelli, Dario and Sparks, James},
  title         = {Accelerating black holes and spinning spindles},
  journal       = {Phys. Rev. D},
  volume        = {104},
  number        = {4},
  pages         = {046007},
  year          = {2021},
  doi           = {10.1103/PhysRevD.104.046007},
  eprint        = {2012.08530},
  archivePrefix = {arXiv},
  primaryClass  = {hep-th}
}

@article{Ferrero:2021ovq,
  author        = {Ferrero, Pietro and Inglese, Matteo and Martelli, Dario and Sparks, James},
  title         = {Multicharge accelerating black holes and spinning spindles},
  journal       = {Phys. Rev. D},
  volume        = {105},
  number        = {12},
  pages         = {126001},
  year          = {2022},
  doi           = {10.1103/PhysRevD.105.126001},
  eprint        = {2109.14625},
  archivePrefix = {arXiv},
  primaryClass  = {hep-th}
}

@article{Guica:2008mu,
  author        = {Guica, Monica and Hartman, Thomas and Song, Wei and Strominger, Andrew},
  title         = {The {Kerr}/{CFT} correspondence},
  journal       = {Phys. Rev. D},
  volume        = {80},
  pages         = {124008},
  year          = {2009},
  doi           = {10.1103/PhysRevD.80.124008},
  eprint        = {0809.4266},
  archivePrefix = {arXiv},
  primaryClass  = {hep-th}
}

@article{Castro:2010fd,
  author        = {Castro, Alejandra and Maloney, Alexander and Strominger, Andrew},
  title         = {Hidden conformal symmetry of the {Kerr} black hole},
  journal       = {Phys. Rev. D},
  volume        = {82},
  pages         = {024008},
  year          = {2010},
  doi           = {10.1103/PhysRevD.82.024008},
  eprint        = {1004.0996},
  archivePrefix = {arXiv},
  primaryClass  = {hep-th}
}

@article{Hosseini:2018dob,
  author        = {Hosseini, Seyed Morteza and Hristov, Kiril and Zaffaroni, Alberto},
  title         = {A note on the entropy of rotating {BPS} {{\(\mathrm{AdS}_{7}\times S^4\)}} black holes},
  journal       = {JHEP},
  volume        = {05},
  pages         = {121},
  year          = {2018},
  doi           = {10.1007/JHEP05(2018)121},
  eprint        = {1803.07568},
  archivePrefix = {arXiv},
  primaryClass  = {hep-th},
  reportNumber  = {IPMU18-0049}
}

@article{Choi:2018fdc,
  author        = {Choi, Sunjin and Hwang, Chiung and Kim, Seok and Nahmgoong, June},
  title         = {Entropy functions of {BPS} black holes in {{\(\mathrm{AdS}_{4}\)}} and {{\(\mathrm{AdS}_{6}\)}}},
  journal       = {J. Korean Phys. Soc.},
  volume        = {76},
  number        = {2},
  pages         = {101--108},
  year          = {2020},
  doi           = {10.3938/jkps.76.101},
  eprint        = {1811.02158},
  archivePrefix = {arXiv},
  primaryClass  = {hep-th},
  reportNumber  = {SNUTP18-006, KIAS-P18076}
}

@article{Cabo-Bizet:2018ehj,
  author        = {Cabo-Bizet, Alejandro and Cassani, Davide and Martelli, Dario and Murthy, Sameer},
  title         = {Microscopic origin of the {Bekenstein}--{Hawking} entropy of supersymmetric {{\(\mathrm{AdS}_{5}\)}} black holes},
  journal       = {JHEP},
  volume        = {10},
  pages         = {062},
  year          = {2019},
  doi           = {10.1007/JHEP10(2019)062},
  eprint        = {1810.11442},
  archivePrefix = {arXiv},
  primaryClass  = {hep-th}
}

@article{Choi:2018hmj,
  author        = {Choi, Sunjin and Kim, Joonho and Kim, Seok and Nahmgoong, June},
  title         = {Large {AdS} black holes from {QFT}},
  year          = {2018},
  month         = {10},
  eprint        = {1810.12067},
  archivePrefix = {arXiv},
  primaryClass  = {hep-th},
  reportNumber  = {SNUTP18-005, KIAS-P18097}
}

@article{Lee:1990nz,
  author        = {Lee, J. and Wald, Robert M.},
  title         = {Local symmetries and constraints},
  journal       = {J. Math. Phys.},
  volume        = {31},
  pages         = {725--743},
  year          = {1990},
  doi           = {10.1063/1.528801}
}

@article{Wald:1999wa,
  author        = {Wald, Robert M. and Zoupas, Andreas},
  title         = {A general definition of `conserved quantities' in general relativity and other theories of gravity},
  journal       = {Phys. Rev. D},
  volume        = {61},
  pages         = {084027},
  year          = {2000},
  doi           = {10.1103/PhysRevD.61.084027},
  eprint        = {gr-qc/9911095},
  archivePrefix = {arXiv}
}

@article{Haco:2018ske,
  author        = {Haco, Sasha and Hawking, Stephen W. and Perry, Malcolm J. and Strominger, Andrew},
  title         = {Black hole entropy and soft hair},
  journal       = {JHEP},
  volume        = {12},
  pages         = {098},
  year          = {2018},
  doi           = {10.1007/JHEP12(2018)098},
  eprint        = {1810.01847},
  archivePrefix = {arXiv},
  primaryClass  = {hep-th}
}

@article{Haco:2019ggi,
  author        = {Haco, Sasha and Perry, Malcolm J. and Strominger, Andrew},
  title         = {{Kerr}--{Newman} black hole entropy and soft hair},
  year          = {2019},
  month         = {2},
  eprint        = {1902.02247},
  archivePrefix = {arXiv},
  primaryClass  = {hep-th}
}

@article{Perry:2020ndy,
  author        = {Perry, Malcolm and Rodriguez, Maria J.},
  title         = {Central charges for {AdS} black holes},
  journal       = {Class. Quant. Grav.},
  volume        = {39},
  number        = {4},
  pages         = {045009},
  year          = {2022},
  doi           = {10.1088/1361-6382/ac45d8},
  eprint        = {2007.03709},
  archivePrefix = {arXiv},
  primaryClass  = {hep-th}
}

@article{Perry:2022udk,
  author        = {Perry, Malcolm J. and Rodriguez, Maria J.},
  title         = {{CFT} duals of {Kerr}--{Taub}--{NUT} and beyond},
  year          = {2022},
  month         = {5},
  eprint        = {2205.09146},
  archivePrefix = {arXiv},
  primaryClass  = {hep-th}
}

@article{Grant:2008sk,
  author        = {Grant, Lars and Grassi, Pietro A. and Kim, Seok and Minwalla, Shiraz},
  title         = {Comments on {{\(1/16\)}} {BPS} quantum states and classical configurations},
  journal       = {JHEP},
  volume        = {05},
  pages         = {049},
  year          = {2008},
  doi           = {10.1088/1126-6708/2008/05/049},
  eprint        = {0803.4183},
  archivePrefix = {arXiv},
  primaryClass  = {hep-th},
  reportNumber  = {IMPERIAL-TP-08-SK-01}
}

@article{Benini:2015eyy,
  author        = {Benini, Francesco and Hristov, Kiril and Zaffaroni, Alberto},
  title         = {Black hole microstates in {{\(\mathrm{AdS}_{4}\)}} from supersymmetric localization},
  journal       = {JHEP},
  volume        = {05},
  pages         = {054},
  year          = {2016},
  doi           = {10.1007/JHEP05(2016)054},
  eprint        = {1511.04085},
  archivePrefix = {arXiv},
  primaryClass  = {hep-th},
  reportNumber  = {IMPERIAL-TP-2015-FB-03}
}

@article{Aharony:2008ug,
  author        = {Aharony, Ofer and Bergman, Oren and Jafferis, Daniel Louis and Maldacena, Juan},
  title         = {{{\(\mathcal{N}=6\)}} superconformal {Chern}--{Simons}-matter theories, {M2}-branes and their gravity duals},
  journal       = {JHEP},
  volume        = {10},
  pages         = {091},
  year          = {2008},
  doi           = {10.1088/1126-6708/2008/10/091},
  eprint        = {0806.1218},
  archivePrefix = {arXiv},
  primaryClass  = {hep-th},
  reportNumber  = {WIS-12-08-JUN-DPP}
}

@article{Chang:2013fba,
  author        = {Chang, Chi-Ming and Yin, Xi},
  title         = {{{\(1/16\)}} {BPS} states in {{\(\mathcal{N}=4\)}} super-{Yang}--{Mills} theory},
  journal       = {Phys. Rev. D},
  volume        = {88},
  number        = {10},
  pages         = {106005},
  year          = {2013},
  doi           = {10.1103/PhysRevD.88.106005},
  eprint        = {1305.6314},
  archivePrefix = {arXiv},
  primaryClass  = {hep-th}
}

@article{Chong:2005hr,
  author        = {Chong, Z. -W. and Cvetic, Mirjam and Lu, H. and Pope, C. N.},
  title         = {General non-extremal rotating black holes in minimal five-dimensional gauged supergravity},
  journal       = {Phys. Rev. Lett.},
  volume        = {95},
  pages         = {161301},
  year          = {2005},
  doi           = {10.1103/PhysRevLett.95.161301},
  eprint        = {hep-th/0506029},
  archivePrefix = {arXiv},
  reportNumber  = {MIFP-05-13, UPR-1125-T}
}

@article{Chong:2004na,
  author        = {Chong, Z. -W. and Cvetic, Mirjam and Lu, H. and Pope, C. N.},
  title         = {Charged rotating black holes in four-dimensional gauged and ungauged supergravities},
  journal       = {Nucl. Phys. B},
  volume        = {717},
  pages         = {246--271},
  year          = {2005},
  doi           = {10.1016/j.nuclphysb.2005.03.034},
  eprint        = {hep-th/0411045},
  archivePrefix = {arXiv},
  reportNumber  = {MIFP-04-20, UPR-1096-T}
}

@article{Cvetic:2005zi,
  author        = {Cvetic, Mirjam and Gibbons, G. W. and Lu, H. and Pope, C. N.},
  title         = {Rotating black holes in gauged supergravities: Thermodynamics, supersymmetric limits, topological solitons and time machines},
  year          = {2005},
  month         = {4},
  eprint        = {hep-th/0504080},
  archivePrefix = {arXiv},
  reportNumber  = {DAMTP-2005-39, MIFP-05-08, UPR-1114-T}
}

@article{Nian:2020qsk,
  author        = {Nian, Jun and Pando Zayas, Leopoldo A.},
  title         = {Toward an effective {{\(\mathrm{CFT}_{2}\)}} from {{\(\mathcal{N}=4\)}} super {Yang}--{Mills} and aspects of {Hawking} radiation},
  journal       = {JHEP},
  volume        = {07},
  pages         = {120},
  year          = {2020},
  doi           = {10.1007/JHEP07(2020)120},
  eprint        = {2003.02770},
  archivePrefix = {arXiv},
  primaryClass  = {hep-th},
  reportNumber  = {LCTP-20-03}
}

@article{Plebanski:1976gy,
  author        = {Plebanski, J. F. and Demianski, M.},
  title         = {Rotating, charged, and uniformly accelerating mass in general relativity},
  journal       = {Annals Phys.},
  volume        = {98},
  pages         = {98--127},
  year          = {1976},
  doi           = {10.1016/0003-4916(76)90240-2}
}

@article{Chong:2006zx,
  author        = {Chong, Z. W. and Cvetic, Mirjam and Lu, H. and Pope, C. N.},
  title         = {Non-extremal rotating black holes in five-dimensional gauged supergravity},
  journal       = {Phys. Lett. B},
  volume        = {644},
  pages         = {192--197},
  year          = {2007},
  doi           = {10.1016/j.physletb.2006.11.012},
  eprint        = {hep-th/0606213},
  archivePrefix = {arXiv},
  reportNumber  = {MIFP-06-17, UPR-1156-T}
}

@article{Gutowski:2004ez,
  author        = {Gutowski, Jan B. and Reall, Harvey S.},
  title         = {Supersymmetric {{\(\mathrm{AdS}_{5}\)}} black holes},
  journal       = {JHEP},
  volume        = {02},
  pages         = {006},
  year          = {2004},
  doi           = {10.1088/1126-6708/2004/02/006},
  eprint        = {hep-th/0401042},
  archivePrefix = {arXiv},
  reportNumber  = {NSF-KITP-04-02}
}

@article{Choi:2019zpz,
  author        = {Choi, Sunjin and Hwang, Chiung and Kim, Seok},
  title         = {Quantum vortices, {M2}-branes and black holes},
  year          = {2019},
  month         = {8},
  eprint        = {1908.02470},
  archivePrefix = {arXiv},
  primaryClass  = {hep-th},
  reportNumber  = {SNUTP18-003}
}

@article{Nian:2019pxj,
  author        = {Nian, Jun and Pando Zayas, Leopoldo A.},
  title         = {Microscopic entropy of rotating electrically charged {{\(\mathrm{AdS}_{4}\)}} black holes from field theory localization},
  journal       = {JHEP},
  volume        = {03},
  pages         = {081},
  year          = {2020},
  doi           = {10.1007/JHEP03(2020)081},
  eprint        = {1909.07943},
  archivePrefix = {arXiv},
  primaryClass  = {hep-th},
  reportNumber  = {LCTP-19-22}
}

@article{Nian:2025iei,
  author        = {Nian, Jun and Pando Zayas, Leopoldo A. and Zheng, Wenni},
  title         = {Explorations of universality in the entropy and {Hawking} radiation of non-extremal {Kerr} {{\(\mathrm{AdS}_{4}\)}} Black Holes},
  journal       = {JHEP},
  volume        = {04},
  pages         = {054},
  year          = {2026},
  doi           = {10.1007/JHEP04(2026)054},
  eprint        = {2508.05322},
  archivePrefix = {arXiv},
  primaryClass  = {hep-th},
  reportNumber  = {PCFT-24-40, LITP-25-06}
}

@article{Kantor:2019lfo,
  author        = {K\'antor, Gergely and Papageorgakis, Constantinos and Richmond, Paul},
  title         = {{{\(\mathrm{AdS}_{7}\)}} black-hole entropy and {5D} {{\(\mathcal{N}=2\)}} {Yang}--{Mills}},
  journal       = {JHEP},
  volume        = {01},
  pages         = {017},
  year          = {2020},
  doi           = {10.1007/JHEP01(2020)017},
  eprint        = {1907.02923},
  archivePrefix = {arXiv},
  primaryClass  = {hep-th},
  reportNumber  = {QMUL-PH-19-17}
}

@article{Nahmgoong:2019hko,
  author        = {Nahmgoong, June},
  title         = {6d superconformal {Cardy} formulas},
  journal       = {JHEP},
  volume        = {02},
  pages         = {092},
  year          = {2021},
  doi           = {10.1007/JHEP02(2021)092},
  eprint        = {1907.12582},
  archivePrefix = {arXiv},
  primaryClass  = {hep-th},
  reportNumber  = {SNUTP19-003}
}

@article{Hosseini:2019iad,
  author        = {Hosseini, Seyed Morteza and Hristov, Kiril and Zaffaroni, Alberto},
  title         = {Gluing gravitational blocks for {AdS} black holes},
  journal       = {JHEP},
  volume        = {12},
  pages         = {168},
  year          = {2019},
  doi           = {10.1007/JHEP12(2019)168},
  eprint        = {1909.10550},
  archivePrefix = {arXiv},
  primaryClass  = {hep-th},
  reportNumber  = {IPMU19-0132}
}

@article{Benini:2016rke,
  author        = {Benini, Francesco and Hristov, Kiril and Zaffaroni, Alberto},
  title         = {Exact microstate counting for dyonic black holes in {{\(\mathrm{AdS}_{4}\)}}},
  journal       = {Phys. Lett. B},
  volume        = {771},
  pages         = {462--466},
  year          = {2017},
  doi           = {10.1016/j.physletb.2017.05.076},
  eprint        = {1608.07294},
  archivePrefix = {arXiv},
  primaryClass  = {hep-th},
  reportNumber  = {SISSA-41-2016-FISI}
}

@article{Chow:2007ts,
  author        = {Chow, David D. K.},
  title         = {Equal charge black holes and seven dimensional gauged supergravity},
  journal       = {Class. Quant. Grav.},
  volume        = {25},
  pages         = {175010},
  year          = {2008},
  doi           = {10.1088/0264-9381/25/17/175010},
  eprint        = {0711.1975},
  archivePrefix = {arXiv},
  primaryClass  = {hep-th},
  reportNumber  = {DAMTP-2007-110}
}

@article{Chong:2004dy,
  author        = {Chong, Z. -W. and Cvetic, Mirjam and Lu, H. and Pope, C. N.},
  title         = {Non-extremal charged rotating black holes in seven-dimensional gauged supergravity},
  journal       = {Phys. Lett. B},
  volume        = {626},
  pages         = {215--222},
  year          = {2005},
  doi           = {10.1016/j.physletb.2005.07.054},
  eprint        = {hep-th/0412094},
  archivePrefix = {arXiv},
  reportNumber  = {MIFP-04-24, UPR-1100-T}
}

@article{Hosseini:2016cyf,
  author        = {Hosseini, Seyed Morteza and Nedelin, Anton and Zaffaroni, Alberto},
  title         = {The {Cardy} limit of the topologically twisted index and black strings in {{\(\mathrm{AdS}_{5}\)}}},
  journal       = {JHEP},
  volume        = {04},
  pages         = {014},
  year          = {2017},
  doi           = {10.1007/JHEP04(2017)014},
  eprint        = {1611.09374},
  archivePrefix = {arXiv},
  primaryClass  = {hep-th}
}

@article{Cvetic:1999xp,
  author        = {Cvetic, Mirjam and Duff, M. J. and Hoxha, P. and Liu, James T. and Lu, Hong and Lu, J. X. and Martinez-Acosta, R. and Pope, C. N. and Sati, H. and Tran, Tuan A.},
  title         = {Embedding {AdS} black holes in ten-dimensions and eleven-dimensions},
  journal       = {Nucl. Phys. B},
  volume        = {558},
  pages         = {96--126},
  year          = {1999},
  doi           = {10.1016/S0550-3213(99)00419-8},
  eprint        = {hep-th/9903214},
  archivePrefix = {arXiv},
  reportNumber  = {UPR-0840-T, CTP-TAMU-11-99, RU-99-4-B}
}

@article{Gutowski:2004yv,
  author        = {Gutowski, Jan B. and Reall, Harvey S.},
  title         = {General supersymmetric {{\(\mathrm{AdS}_{5}\)}} black holes},
  journal       = {JHEP},
  volume        = {04},
  pages         = {048},
  year          = {2004},
  doi           = {10.1088/1126-6708/2004/04/048},
  eprint        = {hep-th/0401129},
  archivePrefix = {arXiv},
  reportNumber  = {NSF-KITP-04-10}
}

@article{Chong:2005da,
  author        = {Chong, Z. W. and Cvetic, Mirjam and Lu, H. and Pope, C. N.},
  title         = {Five-dimensional gauged supergravity black holes with independent rotation parameters},
  journal       = {Phys. Rev. D},
  volume        = {72},
  pages         = {041901},
  year          = {2005},
  doi           = {10.1103/PhysRevD.72.041901},
  eprint        = {hep-th/0505112},
  archivePrefix = {arXiv},
  reportNumber  = {MIFP-05-11, UPR-1123-T}
}

@article{Kunduri:2006ek,
  author        = {Kunduri, Hari K. and Lucietti, James and Reall, Harvey S.},
  title         = {Supersymmetric multi-charge {{\(\mathrm{AdS}_{5}\)}} black holes},
  journal       = {JHEP},
  volume        = {04},
  pages         = {036},
  year          = {2006},
  doi           = {10.1088/1126-6708/2006/04/036},
  eprint        = {hep-th/0601156},
  archivePrefix = {arXiv},
  reportNumber  = {DAMTP-2006-10}
}

@article{Maldacena:1997re,
  author        = {Maldacena, Juan Martin},
  title         = {The large {{\(N\)}} limit of superconformal field theories and supergravity},
  journal       = {Adv. Theor. Math. Phys.},
  volume        = {2},
  pages         = {231--252},
  year          = {1998},
  doi           = {10.4310/ATMP.1998.v2.n2.a1},
  eprint        = {hep-th/9711200},
  archivePrefix = {arXiv},
  reportNumber  = {HUTP-97-A097, HUTP-98-A097}
}

@article{Ma:2026fourcharge,
  author        = {Ma, Liang and L{\"u}, H.},
  title         = {Four-charge Rotating {AdS} Black Holes in Gauged {STU} Supergravity},
  year          = {2026},
  eprint        = {2611.13074},
  archivePrefix = {arXiv},
  primaryClass  = {hep-th}
}

@article{Cacciatori:2016duality,
  author        = {Cacciatori, Sergio L. and Klemm, Dietmar and Rabbiosi, Marco},
  title         = {Duality invariance in {Fayet}--{Iliopoulos} gauged supergravity},
  journal       = {JHEP},
  volume        = {09},
  pages         = {088},
  year          = {2016},
  doi           = {10.1007/JHEP09(2016)088},
  eprint        = {1606.05160},
  archivePrefix = {arXiv},
  primaryClass  = {hep-th}
}

@article{Hristov:2019dyonic,
  author        = {Hristov, Kiril and Katmadas, Stefanos and Toldo, Chiara},
  title         = {Matter-coupled supersymmetric {Kerr}--{Newman}--{{\(\mathrm{AdS}_{4}\)}} black holes},
  journal       = {Phys. Rev. D},
  volume        = {100},
  pages         = {066016},
  year          = {2019},
  doi           = {10.1103/PhysRevD.100.066016},
  eprint        = {1907.05192},
  archivePrefix = {arXiv},
  primaryClass  = {hep-th}
}

@article{Wu:2026fourcharge,
  author        = {Wu, Di and Wu, Shuang-Qing},
  title         = {General non-extremal rotating charged {AdS} black hole solutions with four independent charge parameters in {{\(D=4\)}}, {STU} gauged supergravity},
  year          = {2026},
  eprint        = {2611.13151},
  archivePrefix = {arXiv},
  primaryClass  = {hep-th}
}

@article{Azizi:2016embedding,
  author        = {Azizi, Arash and Godazgar, Hadi and Godazgar, Mahdi and Pope, C. N.},
  title         = {The Embedding of Gauged {STU} Supergravity in Eleven Dimensions},
  journal       = {Phys. Rev. D},
  volume        = {94},
  pages         = {066003},
  year          = {2016},
  doi           = {10.1103/PhysRevD.94.066003},
  eprint        = {1606.06954},
  archivePrefix = {arXiv}
}

@article{Duff:1999electric,
  author        = {Duff, M. J. and Liu, James T.},
  title         = {Anti-de {Sitter} Black Holes in Gauged {{\(\mathcal{N}=8\)}} Supergravity},
  journal       = {Nucl. Phys. B},
  volume        = {554},
  pages         = {237--253},
  year          = {1999},
  doi           = {10.1016/S0550-3213(99)00299-0},
  eprint        = {hep-th/9901149},
  archivePrefix = {arXiv}
}

@article{Caldarelli:1999bps,
  author        = {Caldarelli, Marco M. and Klemm, Dietmar},
  title         = {Supersymmetry of Anti-de {Sitter} Black Holes},
  journal       = {Nucl. Phys. B},
  volume        = {545},
  pages         = {434--460},
  year          = {1999},
  doi           = {10.1016/S0550-3213(98)00846-3},
  eprint        = {hep-th/9808097},
  archivePrefix = {arXiv}
}

@article{Sabra:1999bps,
  author        = {Sabra, W. A.},
  title         = {Anti-de {Sitter} {BPS} Black Holes in {{\(\mathcal{N}=2\)}} Gauged Supergravity},
  journal       = {Phys. Lett. B},
  volume        = {458},
  pages         = {36--42},
  year          = {1999},
  doi           = {10.1016/S0370-2693(99)00564-X},
  eprint        = {hep-th/9903143},
  archivePrefix = {arXiv}
}

@article{Chamseddine:2000dyonic,
  author        = {Chamseddine, A. H. and Sabra, W. A.},
  title         = {Magnetic and Dyonic Black Holes in {{\(D=4\)}} Gauged Supergravity},
  journal       = {Phys. Lett. B},
  volume        = {485},
  pages         = {301--307},
  year          = {2000},
  doi           = {10.1016/S0370-2693(00)00652-3},
  eprint        = {hep-th/0003213},
  archivePrefix = {arXiv}
}

@article{Cacciatori:2010bps,
  author        = {Cacciatori, Sergio L. and Klemm, Dietmar},
  title         = {Supersymmetric {{\(\mathrm{AdS}_{4}\)}} black holes and attractors},
  journal       = {JHEP},
  volume        = {01},
  pages         = {085},
  year          = {2010},
  doi           = {10.1007/JHEP01(2010)085},
  eprint        = {0911.4926},
  archivePrefix = {arXiv}
}

@article{Hristov:2011static,
  author        = {Hristov, Kiril and Vandoren, Stefan},
  title         = {Static supersymmetric black holes in {{\(\mathrm{AdS}_{4}\)}} with spherical symmetry},
  journal       = {JHEP},
  volume        = {04},
  pages         = {047},
  year          = {2011},
  doi           = {10.1007/JHEP04(2011)047},
  eprint        = {1012.4314},
  archivePrefix = {arXiv}
}

@article{Toldo:2012nonextremal,
  author        = {Toldo, Chiara and Vandoren, Stefan},
  title         = {Static nonextremal {{\(\mathrm{AdS}_{4}\)}} black hole solutions},
  journal       = {JHEP},
  volume        = {09},
  pages         = {048},
  year          = {2012},
  doi           = {10.1007/JHEP09(2012)048},
  eprint        = {1207.3014},
  archivePrefix = {arXiv}
}

@article{Gnecchi:2013flow,
  author        = {Gnecchi, Alessandra and Toldo, Chiara},
  title         = {On the non-{BPS} first order flow in {{\(\mathcal{N}=2\)}} {{\(\mathrm{U}(1)\)}}-gauged Supergravity},
  journal       = {JHEP},
  volume        = {03},
  pages         = {088},
  year          = {2013},
  doi           = {10.1007/JHEP03(2013)088},
  eprint        = {1211.1966},
  archivePrefix = {arXiv}
}

@article{Klemm:2013realI,
  author        = {Klemm, Dietmar and Vaughan, Owen},
  title         = {Nonextremal black holes in gauged supergravity and the real formulation of special geometry},
  journal       = {JHEP},
  volume        = {01},
  pages         = {053},
  year          = {2013},
  doi           = {10.1007/JHEP01(2013)053},
  eprint        = {1207.2679},
  archivePrefix = {arXiv}
}

@article{Klemm:2013realII,
  author        = {Klemm, Dietmar and Vaughan, Owen},
  title         = {Nonextremal black holes in gauged supergravity and the real formulation of special geometry {II}},
  journal       = {Class. Quant. Grav.},
  volume        = {30},
  pages         = {065003},
  year          = {2013},
  doi           = {10.1088/0264-9381/30/6/065003},
  eprint        = {1211.1618},
  archivePrefix = {arXiv}
}

@article{Lu:2013dyonic,
  author        = {L{\"u}, H. and Pang, Y. and Pope, C. N.},
  title         = {{AdS} Dyonic Black Hole and its Thermodynamics},
  journal       = {JHEP},
  volume        = {11},
  pages         = {033},
  year          = {2013},
  doi           = {10.1007/JHEP11(2013)033},
  eprint        = {1307.6243},
  archivePrefix = {arXiv}
}

@article{Chow:2014dyonic,
  author        = {Chow, David D. K. and Comp\`ere, Geoffrey},
  title         = {Dyonic {AdS} black holes in maximal gauged supergravity},
  journal       = {Phys. Rev. D},
  volume        = {89},
  pages         = {065003},
  year          = {2014},
  doi           = {10.1103/PhysRevD.89.065003},
  eprint        = {1311.1204},
  archivePrefix = {arXiv}
}

@article{Halmagyi:2015dyonic,
  author        = {Halmagyi, Nick},
  title         = {Static {BPS} Black Holes in {{\(\mathrm{AdS}_{4}\)}} with General Dyonic Charges},
  journal       = {JHEP},
  volume        = {03},
  pages         = {032},
  year          = {2015},
  doi           = {10.1007/JHEP03(2015)032},
  eprint        = {1408.2831},
  archivePrefix = {arXiv}
}

@article{Chow:2011single,
  author        = {Chow, David D. K.},
  title         = {Single-charge rotating black holes in four-dimensional gauged supergravity},
  journal       = {Class. Quant. Grav.},
  volume        = {28},
  pages         = {032001},
  year          = {2011},
  doi           = {10.1088/0264-9381/28/3/032001},
  eprint        = {1011.2202},
  archivePrefix = {arXiv}
}

@article{Chow:2011two,
  author        = {Chow, David D. K.},
  title         = {Two-charge rotating black holes in four-dimensional gauged supergravity},
  journal       = {Class. Quant. Grav.},
  volume        = {28},
  pages         = {175004},
  year          = {2011},
  doi           = {10.1088/0264-9381/28/17/175004},
  eprint        = {1012.1851},
  archivePrefix = {arXiv}
}

@article{Klemm:2011rotating,
  author        = {Klemm, Dietmar},
  title         = {Rotating {BPS} black holes in matter-coupled {{\(\mathrm{AdS}_{4}\)}} supergravity},
  journal       = {JHEP},
  volume        = {07},
  pages         = {019},
  year          = {2011},
  doi           = {10.1007/JHEP07(2011)019},
  eprint        = {1103.4699},
  archivePrefix = {arXiv}
}

@article{Gnecchi:2014rotating,
  author        = {Gnecchi, Alessandra and Hristov, Kiril and Klemm, Dietmar and Toldo, Chiara and Vaughan, Owen},
  title         = {Rotating black holes in {4d} gauged supergravity},
  journal       = {JHEP},
  volume        = {01},
  pages         = {127},
  year          = {2014},
  doi           = {10.1007/JHEP01(2014)127},
  eprint        = {1311.1795},
  archivePrefix = {arXiv}
}

@article{Hristov:2019attractors,
  author        = {Hristov, Kiril and Katmadas, Stefanos and Toldo, Chiara},
  title         = {Rotating attractors and {BPS} black holes in {{\(\mathrm{AdS}_{4}\)}}},
  journal       = {JHEP},
  volume        = {01},
  pages         = {199},
  year          = {2019},
  doi           = {10.1007/JHEP01(2019)199},
  eprint        = {1811.00292},
  archivePrefix = {arXiv}
}

@article{Halmagyi:2014duality,
  author        = {Halmagyi, Nick and Vanel, Thomas},
  title         = {{AdS} Black Holes from Duality in Gauged Supergravity},
  journal       = {JHEP},
  volume        = {04},
  pages         = {130},
  year          = {2014},
  doi           = {10.1007/JHEP04(2014)130},
  eprint        = {1312.5430},
  archivePrefix = {arXiv}
}

@article{Chow:2014ungauged,
  author        = {Chow, David D. K. and Comp\`ere, Geoffrey},
  title         = {Black holes in {{\(\mathcal{N}=8\)}} supergravity from {{\(\mathrm{SO}(4,4)\)}} hidden symmetries},
  journal       = {Phys. Rev. D},
  volume        = {90},
  pages         = {025029},
  year          = {2014},
  doi           = {10.1103/PhysRevD.90.025029},
  eprint        = {1404.2602},
  archivePrefix = {arXiv}
}

@article{Katmadas:2014static,
  author        = {Katmadas, Stefanos},
  title         = {Static {BPS} black holes in {{\(\mathrm{U}(1)\)}} gauged supergravity},
  journal       = {JHEP},
  volume        = {09},
  pages         = {027},
  year          = {2014},
  doi           = {10.1007/JHEP09(2014)027},
  eprint        = {1405.4901},
  archivePrefix = {arXiv}
}

@article{Griffiths:2006pd,
  author        = {Griffiths, J. B. and Podolsk{\'y}, J.},
  title         = {A new look at the {Pleba{\'n}ski--Demia{\'n}ski} family of solutions},
  journal       = {Int. J. Mod. Phys. D},
  volume        = {15},
  number        = {3},
  pages         = {335--369},
  year          = {2006},
  doi           = {10.1142/S0218271806007742},
  eprint        = {gr-qc/0511091},
  archivePrefix = {arXiv}
}

@article{Cvetic:2004threecharge,
  author        = {Cveti{\v{c}}, Mirjam and L{\"u}, Hong and Pope, C. N.},
  title         = {Charged rotating black holes in five dimensional {{\(\mathrm{U}(1)^3\)}} gauged {{\(\mathcal{N}=2\)}} supergravity},
  journal       = {Phys. Rev. D},
  volume        = {70},
  pages         = {081502},
  year          = {2004},
  doi           = {10.1103/PhysRevD.70.081502},
  eprint        = {hep-th/0407058},
  archivePrefix = {arXiv}
}

@article{Wu:2012general5d,
  author        = {Wu, Shuang-Qing},
  title         = {General nonextremal rotating charged {AdS} black holes in five-dimensional {{\(\mathrm{U}(1)^3\)}} gauged supergravity: A simple construction method},
  journal       = {Phys. Lett. B},
  volume        = {707},
  pages         = {286--291},
  year          = {2012},
  doi           = {10.1016/j.physletb.2011.12.031},
  eprint        = {1108.4159},
  archivePrefix = {arXiv},
  primaryClass  = {hep-th}
}

@article{BlazquezSalcedo:2017magnetized,
  author        = {Bl{\'a}zquez-Salcedo, Jose Luis and Kunz, Jutta and Navarro-L{\'e}rida, Francisco and Radu, Eugen},
  title         = {{{\(\mathrm{AdS}_{5}\)}} magnetized solutions in minimal gauged supergravity},
  journal       = {Phys. Lett. B},
  volume        = {771},
  pages         = {52--58},
  year          = {2017},
  doi           = {10.1016/j.physletb.2017.05.014},
  eprint        = {1703.04163},
  archivePrefix = {arXiv},
  primaryClass  = {gr-qc}
}

@article{Behrndt:1999static5d,
  author        = {Behrndt, Klaus and Cveti{\v{c}}, Mirjam and Sabra, W. A.},
  title         = {Non-Extreme Black Holes of Five Dimensional {{\(\mathcal{N}=2\)}} {AdS} Supergravity},
  journal       = {Nucl. Phys. B},
  volume        = {553},
  pages         = {317--332},
  year          = {1999},
  doi           = {10.1016/S0550-3213(99)00243-6},
  eprint        = {hep-th/9810227},
  archivePrefix = {arXiv}
}

@article{Iyer:1994noether,
  author        = {Iyer, Vivek and Wald, Robert M.},
  title         = {Some properties of {Noether} charge and a proposal for dynamical black hole entropy},
  journal       = {Phys. Rev. D},
  volume        = {50},
  pages         = {846--864},
  year          = {1994},
  doi           = {10.1103/PhysRevD.50.846},
  eprint        = {gr-qc/9403028},
  archivePrefix = {arXiv}
}

@article{Appels:2017conical,
  author        = {Appels, Michael and Gregory, Ruth and Kubiznak, David},
  title         = {Black hole thermodynamics with conical defects},
  journal       = {JHEP},
  volume        = {05},
  pages         = {116},
  year          = {2017},
  doi           = {10.1007/JHEP05(2017)116},
  eprint        = {1702.00490},
  archivePrefix = {arXiv},
  primaryClass  = {hep-th}
}

@article{Benini:2018mlo,
  author        = {Benini, Francesco and Milan, Elisa},
  title         = {A {Bethe} Ansatz type formula for the superconformal index},
  journal       = {Commun. Math. Phys.},
  volume        = {376},
  number        = {2},
  pages         = {1413--1440},
  year          = {2020},
  doi           = {10.1007/s00220-019-03679-y},
  eprint        = {1811.04107},
  archivePrefix = {arXiv},
  primaryClass  = {hep-th},
  reportNumber  = {SISSA 46/2018/FISI}
}

@article{Cabo-Bizet:2019eaf,
  author        = {Cabo-Bizet, Alejandro and Murthy, Sameer},
  title         = {Supersymmetric phases of 4d {{\(\mathcal{N}=4\)}} {SYM} at large {{\(N\)}}},
  journal       = {JHEP},
  volume        = {09},
  pages         = {184},
  year          = {2020},
  doi           = {10.1007/JHEP09(2020)184},
  eprint        = {1909.09597},
  archivePrefix = {arXiv},
  primaryClass  = {hep-th}
}

@article{Murthy:2020scj,
  author        = {Murthy, Sameer},
  title         = {Growth of the {{\(1/16\)}}-{BPS} index in 4d {{\(\mathcal{N}=4\)}} supersymmetric {Yang}--{Mills} theory},
  journal       = {Phys. Rev. D},
  volume        = {105},
  number        = {2},
  pages         = {L021903},
  year          = {2022},
  doi           = {10.1103/PhysRevD.105.L021903},
  eprint        = {2005.10843},
  archivePrefix = {arXiv},
  primaryClass  = {hep-th}
}

@article{Agarwal:2020zwm,
  author        = {Agarwal, Prarit and Choi, Sunjin and Kim, Joonho and Kim, Seok and Nahmgoong, June},
  title         = {{AdS} black holes and finite {{\(N\)}} indices},
  journal       = {Phys. Rev. D},
  volume        = {103},
  number        = {12},
  pages         = {126006},
  year          = {2021},
  doi           = {10.1103/PhysRevD.103.126006},
  eprint        = {2005.11240},
  archivePrefix = {arXiv},
  primaryClass  = {hep-th},
  reportNumber  = {QMUL-PH-20-11, SNUTP20-001, KIAS-P20020}
}

@article{Chang:2024zqi,
  author        = {Chang, Chi-Ming and Lin, Ying-Hsuan},
  title         = {Holographic covering and the fortuity of black holes},
  year          = {2024},
  month         = {2},
  eprint        = {2402.10129},
  archivePrefix = {arXiv},
  primaryClass  = {hep-th}
}

@article{Kapustin:2011jm,
  author        = {Kapustin, Anton and Willett, Brian},
  title         = {Generalized Superconformal Index for Three Dimensional Field Theories},
  year          = {2011},
  month         = {6},
  eprint        = {1106.2484},
  archivePrefix = {arXiv},
  primaryClass  = {hep-th}
}

@article{Liu:1999ads7,
  author        = {Liu, James T. and Minasian, Ruben},
  title         = {Black holes and membranes in {{\(\mathrm{AdS}_{7}\)}}},
  journal       = {Phys. Lett. B},
  volume        = {457},
  pages         = {39--46},
  year          = {1999},
  doi           = {10.1016/S0370-2693(99)00500-6},
  eprint        = {hep-th/9903269},
  archivePrefix = {arXiv}
}

@article{Wu:2011ads7,
  author        = {Wu, Shuang-Qing},
  title         = {Two-charged non-extremal rotating black holes in seven-dimensional gauged supergravity: The single-rotation case},
  journal       = {Phys. Lett. B},
  volume        = {705},
  pages         = {383--387},
  year          = {2011},
  doi           = {10.1016/j.physletb.2011.10.026},
  eprint        = {1108.4158},
  archivePrefix = {arXiv},
  primaryClass  = {hep-th}
}

@article{Chow:2011singleRotation,
  author        = {Chow, David D. K.},
  title         = {Single-rotation two-charge black holes in gauged supergravity},
  year          = {2011},
  eprint        = {1108.5139},
  archivePrefix = {arXiv},
  primaryClass  = {hep-th},
  reportNumber  = {MIFPA-2011-37}
}

@article{Bobev:2023ads7,
  author        = {Bobev, Nikolay and David, Marina and Hong, Junho and Mouland, Rishi},
  title         = {{{\(\mathrm{AdS}_{7}\)}} black holes from rotating {M5}-branes},
  journal       = {JHEP},
  volume        = {09},
  pages         = {143},
  year          = {2023},
  doi           = {10.1007/JHEP09(2023)143},
  eprint        = {2307.06364},
  archivePrefix = {arXiv},
  primaryClass  = {hep-th},
  note          = {Erratum: JHEP 09 (2023) 198}
}

@article{Bobev:2025ads7holography,
  author        = {Bobev, Nikolay and David, Marina and Dimitrov, Vasil and Hong, Junho},
  title         = {{{\(\mathrm{AdS}_{7}\)}} black holes and holography},
  journal       = {JHEP},
  volume        = {11},
  pages         = {159},
  year          = {2025},
  doi           = {10.1007/JHEP11(2025)159},
  eprint        = {2509.02833},
  archivePrefix = {arXiv},
  primaryClass  = {hep-th}
}

@article{Sen:2005entropy,
  author        = {Sen, Ashoke},
  title         = {Black hole entropy function and the attractor mechanism in higher derivative gravity},
  journal       = {JHEP},
  volume        = {09},
  pages         = {038},
  year          = {2005},
  doi           = {10.1088/1126-6708/2005/09/038},
  eprint        = {hep-th/0506177},
  archivePrefix = {arXiv}
}

@article{Sen:2008quantum,
  author        = {Sen, Ashoke},
  title         = {Quantum entropy function from {{\(\mathrm{AdS}_{2}\)}}/{{\(\mathrm{CFT}_{1}\)}} correspondence},
  journal       = {Int. J. Mod. Phys. A},
  volume        = {24},
  pages         = {4225--4244},
  year          = {2009},
  doi           = {10.1142/S0217751X09045893},
  eprint        = {0809.3304},
  archivePrefix = {arXiv},
  primaryClass  = {hep-th}
}

@article{Marolf:2006boundary,
  author        = {Marolf, Donald and Ross, Simon F.},
  title         = {Boundary Conditions and Dualities: Vector Fields in {AdS/CFT}},
  journal       = {JHEP},
  volume        = {11},
  pages         = {085},
  year          = {2006},
  doi           = {10.1088/1126-6708/2006/11/085},
  eprint        = {hep-th/0606113},
  archivePrefix = {arXiv}
}

\end{document}